\documentclass{emulateapj}

\usepackage{bm}

\newcommand{\novaper} {GRO~J0422+32}
\newcommand{\novamus} {GS~1124--684}

\newcommand{\novamon} {A~0620--00}

\newcommand{\novavel} {GRS~1009--45}

\newcommand{\novaophb}{GRS~1716--249}
\newcommand{\novavula} {GS~2000+25}
\newcommand{\novavulb} {XTE~J1859+226}

\newcommand{\ANS}  {\textit{ANS}}
\newcommand{\ASCA} {\textit{ASCA}}
\newcommand{\ROSAT}{\textit{ROSAT}}
\newcommand{\HST}  {\textit{HST}}

\newcommand{\RXTE} {\textit{RXTE}}

\newcommand{\IUE}  {\textit{IUE}}
\newcommand{\SAX}  {\textit{SAX}}
\newcommand{\Ginga}  {\textit{Ginga}}
\newcommand{\EBV}{E(B-V)}
\newcommand{\ebv}{E(B-V)}
\newcommand{\Rv} {R_{\rm V}}
\newcommand{\Av} {A_{\rm V}}
\slugcomment{Submitted to the Astrophysical Journal}

\shorttitle{Outburst SEDs of Black Hole X-ray Transients}
\shortauthors{Hynes}

\begin{document}


\title{The Optical and Ultraviolet Spectral Energy Distributions of Short Period Black Hole
X-ray Transients in Outburst}

\author{R. I. Hynes\altaffilmark{1}}
\affil{McDonald Observatory and 
Astronomy Department, The University of Texas at Austin, 
1 University Station C1400, Austin, Texas 78712, USA} 
\affil{
Department of Physics and
  Astronomy, Louisiana State University, Baton Rouge, 
  Louisiana 70803, USA}
\altaffiltext{1}{rih@phys.lsu.edu}

\begin{abstract}
We compile optical and UV spectra of a sample of `typical' short
period black hole X-ray transients in outburst.  We also survey
determinations of interstellar extinction and distance in order to
deredden spectra and compare absolute fluxes.  Hence we perform a
comparative study of the broad-band spectral energy distributions
(SED).  We find that given such a homogeneous sample of typical
sources, the optical SEDs form a relatively uniform set, all
exhibiting quasi-power-law spectra with $F_\nu \propto \nu^{\alpha}$,
where $0.5\la\alpha\la1.5$ (steeper than the canonical $\nu^{1/3}$
disk spectrum).  All become flatter in the UV, although there is more
diversity here.  The SEDs studied can be broadly divided into two
optical-UV spectral states.  The UV-hard spectra, e.g.\ \novamon\ and
\novamus, continue to rise in the far-UV.  The UV-soft spectra, e.g.\
\novaper, drop off.  \novavulb\ evolved from UV-soft to UV-hard as it
decayed indicating that this effect is a real difference, not a
dereddening artifact.  All of the spectra can be fitted by a
generalized black body disk model with two forms of heating, resulting
in the two states.  The UV-soft state is consistent with a disk
illuminated by a central point source, with irradiative heating
dominating over viscous.  The UV-hard state is well described by a
viscously heated disk, although this requires very high mass flow
rates in the case of \novamus.  Alternatively, a UV-hard spectrum can
be produced if the disk is illuminated by a vertically extended X-ray
source such as a central scattering corona or jet.  Since scattering
is assumed by some numerical simulations, it is worth emphasizing that
where illumination comes from (non-local) scattering high above the
disk, we generically expect a steeper radial dependence of X-ray
heating ($F\propto R^{-3}$) than is usually assumed; it is this steep
dependence that leads to the UV-hard spectrum.
\end{abstract}


\section{Introduction}
Black hole X-ray transients (BHXRTs), also known as X-ray novae and
soft X-ray transients, are low-mass X-ray binaries (LMXBs) which
undergo dramatic X-ray, optical, and radio outbursts, separated by
years, decades or even longer periods of quiescence
(\citealt{Tanaka:1996a};\citealt{Cherepashchuk:2000a}).  The class is
diverse, with orbital periods of 4\,hrs to a few days to a month.
Outburst behavior is similarly varied.  Among these, however, are a
sub-group which show relatively consistent behavior and these are
thought of as the `typical' members of the class.  They are often
compared to the prototypical BHXRT, A\,0620--00 (see
\citealt{Kuulkers:1998a} for a review).  In such a typical outburst,
the X-ray emission is usually (but not always, see
\citealt{Brocksopp:2004a}) dominated by thermal emission from the hot
inner accretion disk, and UV and optical emission is thought to be
produced by reprocessing of X-rays by the outer disk.

A number of attempts have been made to characterize the spectral
energy distributions (SEDs) of these BHXRTs in outburst.
\citet{Cheng:1992a} and \citet{Hynes:2002a} attempted to use simple
disk spectral models, and other authors have used a variety of black
bodies and power-laws.  These mostly only consider a single source at
a time, use a variety of extinction curves and reddening determination
methods, and do not use comparable spectral models.  Consequently it
is difficult to compare the data on different sources and attempt to
draw broad conclusions about the population as a whole.  The goal of
this paper is to bring together all of the useful data on a suitable
homogeneous sample of sources, with recalibrations where appropriate
for the satellite data.  It is then possible to deredden the spectra
consistently and draw direct empirical conclusions about the observed
SEDs.  We can also go on to compare the complete sample with simple
models in a consistent way.  Finally we can hope that this compilation
will provide a reference set as more sophisticated disk spectral
models are developed.

\begin{deluxetable*}{llrccc}
\tablecaption{Sources included in the short period sample considered.\label{SourceTable}}
\startdata
\hline
\noalign{\smallskip}
Object & Outburst peak & $P_{\rm orb}$ & Adopted                 & Dust map                & Distance\tablenotemark{a} \\
       & (approx.)     & (hrs)         & $\ebv$\tablenotemark{a} & $\ebv$\tablenotemark{b} & (kpc)\\
\noalign{\smallskip}
\novamon\  = V616~Mon\tablenotemark & 1975 Aug 13  (MJD 42637)&      7.8 &
$0.35\pm0.02$ & 0.49 & $1.16\pm0.11$ \\
\novamus\  = GU~Mus\tablenotemark   & 1991 Jan 15  (MJD 48271)&     10.4 &
$0.30\pm0.06$ & 0.37 & $5.89\pm0.26$ \\
\novaper\  = V518~Per\tablenotemark & 1992 Aug 15  (MJD 48848)&      5.1 &
$0.35\pm0.10$ & 0.36 & $2.49\pm0.30$ \\
\novavel\  = MM~Vel    & 1993 Sep 15 (MJD 49245)&      6.9 &
$0.20\pm0.05$ & 0.19 & $3.82\pm0.27$  \\
\novaophb\ = V2293~Oph & 1993 Sep 30 (MJD 49260)&$\la$14.7\tablenotemark{c} &
$0.90\pm0.20$ & 1.08 & ($0.24\pm0.04$) \\
\novavulb\ = V406~Vul\tablenotemark & 1999 Oct 16  (MJD 51467)&      9.1 &
$0.58\pm0.12$ & 0.49 & ($8\pm3$) \\
\noalign{\smallskip}
\hline
\enddata
\tablenotetext{a}{Sources for reddening and distance estimates are
  discussed in the text.}
\tablenotetext{b}{From \citet{Schlegel:1998a}}
\tablenotetext{c}{\novaophb\ does not have a confirmed period, but
possible superhumps seen in outburst with a 14.7\,hr period suggest
that the orbital period is similar to this.}
\end{deluxetable*}

Studies of these broad-band SEDs are important as they can inform us
about the temperature structure within the binary.  They are
especially useful for the typical BHXRT systems considered, as these
mostly have very small mass ratios, hence the emission from
reprocessing on the companion star is weak, and the systems so far
discovered have intermediate or low inclinations, so disk self
obscuration and limb darkening should not be large effects.  In these
cases the comparison between observed SED and disk temperature
structure should be relatively direct.  One must of course assume an
emergent spectrum for the disk.  At present, black bodies are probably
as good an approximation as any for optically thick disks in X-ray
binaries, but we can anticipate models including a more rigorous
treatment of the disk atmosphere in the future.  The observed SED
essentially can tell us the distribution of emitting area as a
function of temperature.  Given assumptions such as axisymmetry and a
monotonic dependence of temperature on radius then one can map this
area--temperature relation into the radial temperature distribution,
but the more generalized statement is useful in considering more
complex geometries which could be imagined, such as disks which are
warped or exhibit spiral waves.  In this more general case, given a
three-dimensional disk geometry, a model for the irradiation geometry
(e.g.\ an isotropically emitting point source, or perhaps a flat
central disk), and some method to calculate emergent spectra (at its
simplest, a black body), one can test specific models against the
observed spectra, and falsify some combinations.  There is certainly
more than enough uncertainty about the temperature structure of
irradiated disks to warrant seeking further observational constraints.
For example, theoretical work suggests that disks should be
self-shielding (\citealt{Meyer:1982a}; \citealt{Tuchman:1990a};
\citealt{Dubus:1999a}), and hence that irradiation requires scattering
of X-rays or warping of the disk \citep{Dubus:1999a}.  These two cases
may produce very different temperature structures.  An understanding
of irradiation geometry has immediate application in providing input
to numerical models of BHXRT outbursts, as these models have to make
some assumptions about the strength, and radial dependence, of X-ray
heating (e.g.\ \citealt{Kim:1999a}; \citealt{Esin:2000a};
\citealt{Dubus:2001a}; \citealt{Truss:2002a}).

We begin in Section~\ref{SelectionSection} by defining our sample of
BHXRTs.  In Section~\ref{ReddeningSection} we review and evaluate
reddening determinations for these sources, in particular
incorporating a reanalysis of 2175\,\AA\ bump data for some
sources. This section also discusses the importance of the extinction
curve adopted.  Section~\ref{DistanceSection} discusses corresponding
distance estimates.  We then proceed to review the database of spectra
available based on satellite observations (mainly UV) in
Section~\ref{SatelliteSection}, and ground-based observations
(optical) in Section~\ref{OpticalSection}.  Having collated our sample
spectra, we then compare the optical spectra in
Section~\ref{OpticalSEDSection} and the broader-band optical-UV
spectra in Section~\ref{UVSEDSection}.  We next take an interlude to
evaluate the effect of reddening uncertainties upon the spectra in
Section~\ref{DereddenVarSection} before comparing the spectra with
disk models in Section~\ref{ModelSection}.  In
Section~\ref{DiscussionSection} we compare our results with other
studies and with theoretical work.  Finally in
Section~\ref{ConclusionSection} we summarize our conclusions and
consider where progress can most effectively be made in the future.
%
%
\section{Source Selection}
\label{SelectionSection}

We restrict our sample of sources in order to obtain a relatively
homogeneous set of spectra.  Our selection criteria are as follows.

Firstly we consider only sources which are believed to harbor a black
hole.  In most cases this selection is based on a dynamical mass
estimate, but where this is not available, more subjective criteria
are used.  Any source exhibiting type I X-ray bursts is obviously
excluded.

Secondly we exclude those sources with an orbital period greater than
one day.  Most long period objects exhibit unusual outbursts, several
have companions luminous enough to contribute even during outburst and
the large disks expected in these systems may well behave rather
differently to those in the short period systems.

Finally we only include high luminosity systems showing a `canonical'
lightcurve, i.e.\ one with a fast rise and quasi-exponential decay.
Specifically we exclude XTE~J1118+480; this seems appropriate as this
source exhibited an unusually faint outburst and the optical light
probably includes significant non-disk emission, likely of a
synchrotron origin (e.g.\ \citealt{Hynes:2000a}).  For the sources
included, only spectra from the exponential decay phase of the
lightcurve are used, i.e.\ we exclude mini-outburst spectra.

An additional practical criterion is of course that there must exist
flux-calibrated spectra.  Based on these criteria we obtain a primary
sample of sources with relatively complete optical and UV coverage and
a secondary sample with only optical spectra.  Our primary sample
includes \novamon, \novamus, \novaper, and XTE\,J1859+226.  Our
secondary sample adds \novavel, and \novaophb.  We initially included
\novavula\ in the secondary sample, as a fluxed spectrum was presented
by \citet{Charles:1991a} and it otherwise meets our selection
criteria.  When we came to estimate the reddening, however, it was
clear that it was too large, and too uncertain, for a meaningful
derivation of the intrinsic spectral energy distribution, so we have
not attempted an analysis of this spectrum.  For all of the other
targets selected $\ebv < 1$ and is relatively well determined.  All of
these objects also lie at least $5^{\circ}$ from the Galactic plane.

For the sources considered, the optical and UV emission should be
dominated by the disk, although the IR may in some cases include a
synchrotron component (e.g.\ \citealt{Brocksopp:2002a};
\citealt{Corbel:2002a}), so we do not include IR coverage.
%
%
\section{Reddening estimates}
\label{ReddeningSection}
\subsection{Background}
Since most BHXRTs lie at significant distances, they are typically
rather reddened objects, hence a reliable correction for interstellar
reddening is crucial if we are to accurately reconstruct the spectral
energy distribution.  This correction depends upon two things: the
determination of the reddening value, typically parameterized in
$\ebv$, and an extinction curve specifying the amount of extinction as
a function of wavelength.  For a review of the problem see
\citet{Fitzpatrick:1999a}.

A number of methods are available to estimate the reddening, but they
vary considerably in their reliability.  The most direct measures are
not necessarily the best; these include the optical color (e.g.\
$B-V$) or emission line ratios.  These do directly measure the
differential extinction as a function of wavelength, but their
interpretation is model dependent.  Less direct methods can be more
independent of assumptions about the target, but instead involve
assumed properties for the line-of-sight material, for example the gas
to dust ratio, the abundance of Na\,{\sc I}, or the strength of the
2175\,\AA\ absorption feature relative to the broader structure in the
extinction curve.

Several approaches also exist for choosing an extinction curve, as
discussed by \citet{Hynes:2002a}.  As we typically have no independent
information on the properties of the interstellar medium toward a
BHXRT, we will adopt the \citet{Fitzpatrick:1999a} $R_V=3.1$
extinction curve as this should be the `best' current Galactic average
curve.  In analyzing our results we must then remember that our
uncertainties depend not only on the uncertain amount of extinction
but the uncertain shape of the extinction curve.

\subsection{Methods}
\subsubsection{Interstellar spectral features}
Our preferred method to estimate the amount of extinction is to
measure the strength of the 2175\,\AA\ interstellar absorption
feature.  In fitting the feature we must assume some underlying
spectral shape, either fixed or with some free parameters, and then
adjust $E(B-V)$ to fit the data.  It should be borne in mind that the
strength of the feature is one of the variable characteristics of the
UV extinction curve; hence the correlation between the 2175\,\AA\
feature and $\ebv$ is not perfect, and the intrinsic variance in their
relative values introduces an unavoidable 20\,percent error in
reddenings determined in this way \citep{Fitzpatrick:1999a}.  This
uncertainty can only be avoided if the relative strength of the
feature for the line of sight is known independently from
nearby stars, which is only the case for one of our sources,
\novamon\ \citep{Wu:1983a}.  In spite of these limitations, the method
is generally more precise than the alternatives.

The equivalent width of the Na\,D interstellar doublet can in
principle be used to estimate the reddening, and calibrations exist
for this purpose.  The problem is that {\em several} calibrations
exist and these do not agree.  \citet{Barbon:1990a} derive
$\EBV=0.25\times EW$ from observations of supernovae in external
galaxies.  \citet{DellaValle:1993a} on the other hand obtained
$\EBV=0.61\times EW - 0.08$ by analyzing the data of
\citet{Cohen:1975a} on southern supergiants.  Furthermore, as noted by
\citet{Munari:1997a}, the Na\,D lines are not in general sensitive to
$\EBV \geq 0.5$ due to saturation, so this method can underestimate
the reddening; whether it does or not depends on the detailed
substructure of the lines which is not usually resolved.  Some caution
should therefore be exercised in applying this method.

In addition to the Na\,D lines, relations between the equivalent
widths of diffuse interstellar bands and $\EBV$ are commonly used.
\citet{Herbig:1975a} performed an extensive study and tabulated
relations for 17 interstellar bands in the 4400--6850\,\AA\ range.
The correlation between different lines was found to be very good but
there were significant regional effects.

\subsubsection{X-ray absorption}

Another approach taken is to adopt scaling ratios between interstellar
gas and dust densities, usually expressed in a regression relation
between $N_{\rm H}$ and $\Av$ or $\ebv$.  This method is plagued with
uncertainty, both due to intrinsic variance in this ratio, and due to
different measures of $N_{\rm H}$; it may be measured directly though
UV Ly\,$\alpha$ absorption (e.g.\ \citealt{Savage:1972a};
\citealt{Bohlin:1978a}) or indirectly through soft X-ray absorption
which also includes metallic absorption (e.g.\
\citealt{Gorenstein:1975a}; \citealt{Predehl:1995a}); for the latter
it is necessary to assume abundances.  We will adopt the relation of
\citet{Bohlin:1978a}: $N_{\rm H} = 5.8\times10^{21}$\,cm$^{-2} \times
\ebv$, as this relation is relatively direct and conveniently falls
intermediate between the other estimates referred to.  These authors
note that individual sources typically scatter by 30\,percent around
this line, and adopting this uncertainty encompasses the other average
relationships as well.  An additional concern for measurements of
N$_{\rm H}$ is the possible presence of local absorption intrinsic to
the source.  The local absorbing material will probably be too hot to
include a dusty component, so will not contribute to the optical/UV
absorption.

\subsubsection{Outburst colors and line ratios}

Some authors assume a typical intrinsic $(B-V)$ color in outburst, and
hence derive a direct estimate of $\ebv$.  For our purposes, however,
this assumption is extremely dangerous; to use such estimates without
independent corroboration would lead to a circular logic.  We would be
assuming that the intrinsic spectra of all BHXRTs are the same to
determine their reddening and hence derive their intrinsic spectrum.

A direct measure of the reddening of the spectrum can alternatively be
obtained from the Balmer decrement, by assuming line ratios
appropriate for Case B recombination (e.g.\
\citealt{Osterbrock:1989a}).  This approach is obviously dependent
upon the assumption that Case B line ratios are appropriate, and so is
also not a very secure method.

\subsubsection{Extinction maps}

A final technique, of more use for extragalactic work, is to use maps
of Galactic dust, for example \citet{Schlegel:1998a}.  The problem
with applying this technique to Galactic objects is that it only
yields the total amount of reddening along that line of sight, some of
which may be behind the target.  This technique is thus mainly useful
as an upper limit, although some sources such as \novaper\ are
believed to lie above the dusty component of the Galactic disk.  An
additional concern is that these maps do not resolve fine extinction
structure and so are only reliable for sources above or below the
Galactic plane; \citet{Schlegel:1998a} recommend only trusting
$\left|b\right| \ga 5^{\circ}$.  Fortunately, the nature of our sample
is such that highly reddened objects in the plane have been excluded,
and all of our targets happen to lie outside of this band.  We
therefore apply this technique, remembering that it may include
background extinction.  In most cases the reddening estimated in this
way is comparable to, or somewhat larger than, other methods so it
does seem to be of use, at least for high latitude X-ray binaries.

\subsection{Objects}

\subsubsection{\novamon}
\citet{Wu:1976a} derived $\ebv = 0.39\pm0.02$ for \novamon\ from
fitting \ANS\ UV data through 2175\,\AA\ feature.  Subsequently,
\citet{Wu:1983a} used spectra from 25 nearby early type stars to
derive a line-of-sight extinction curve for \novamon.  It was found
that the 2175\,\AA\ feature was rather stronger along this
line-of-sight than average, and hence they derived a revised lower
estimate of $\ebv=0.35$.  The uncertainty in this estimate will be
similar to the earlier measurement; since they used a line-of-sight
extinction curve there is no need to account for a 20\,percent error
due to uncertainty in the intrinsic 2175\,\AA\ strength.
\citet{Oke:1977a} estimated $\ebv\sim0.44$, with a large uncertainty,
from the Na\,D lines.  They note, however, that the DIBs seem
unusually strong for the reddening derived.  \citet{Whelan:1977a}
collated the DIB measurements and confirmed this conclusion.  They
emphasize, however, that the strength of DIBs as a function of $\ebv$
does vary with Galactic longitude, and that \novamon\ lies within
$5^{\circ}$ of the region where the DIBs are relatively strongest.
Using appropriate relations for this longitude they derive
$\ebv=0.25\pm0.05$ or $0.38\pm0.07$, depending on the relation used.
The reddening derived from optical interstellar features is therefore
broadly consistent with that obtained from the UV.  Both are lower
than the value of 0.49 derived from dust maps \citep{Schlegel:1998a}
but \novamon\ is the closest of our objects and background extinction
is expected.  We adopt the \ANS\ value of $0.35\pm0.02$ for the
remainder of this work.
\subsubsection{\novamus}
\novamus\ was among the best studied in the UV with a series of \IUE\
observations \citep{Shrader:1993a} early in the outburst and a later,
higher signal-to-noise \HST\ one \citep{Cheng:1992a}.  These permit
fitting of the 2175\,\AA\ feature to derive $\EBV$.
\citet{Shrader:1993a} derive $\EBV=0.30\pm0.05$ using this method and
\citet{Cheng:1992a} obtain $\EBV=0.287\pm0.004$ from considerably
higher signal-to-noise data.  Both use the \citet{Seaton:1979a}
extinction law.  Repeating the analysis of the \HST\ data using the
\citet{Fitzpatrick:1999a} extinction curve, and a more realistic
estimate of the accuracy of the method, we derive $\ebv=0.30\pm0.06$
from these data in Section~\ref{HSTSection}, consistent with both of
the earlier estimates.  From interstellar absorption lines,
\citet{DellaValle:1991a} obtain two values: 0.3 from applying the
calibration of \citet{Barbon:1990a} to the Na\,D lines (with a
measured equivalent width of 1.4\,\AA) and 0.35 using the calibration
of \citet{Herbig:1975a} on the 5778, 5780, 5797\,\AA\ absorption
bands.  These are consistent with the UV determination.  The hydrogen
column density has been estimated from \Ginga\ data alone as about
$1.6\times10^{21}$\,cm$^{-2}$ \citep{Ebisawa:1994a}, and from \Ginga\
and \ROSAT\ spectra at $2.2\times10^{21}$\,cm$^{-2}$
\citep{Greiner:1994a}.  These imply reddenings of $\ebv\sim0.3$ and
0.4 respectively, which are just about consistent with optical
estimates given the uncertainty in the relation between the two.  For
the remainder of this work we will adopt our revised estimate based on
the 2175\,\AA\ feature: $\ebv=0.30\pm0.06$.  As for \novamon, this is
slightly lower than was derived from dust maps, 0.37.
\subsubsection{\novaper}
The 2175\,\AA\ feature in \novaper\ has been measured by
\citet{Shrader:1994a} using \IUE\ data.  They obtain
$\EBV=0.40\pm0.06$.  We have repeated the analysis using a more recent
extinction curve and obtain a range of values of 0.29--0.48, depending
on the spectral range modeled and the underlying spectrum which is
assumed (see Section~\ref{IUESection}).  Many authors have also
attempted to derive a reddening from interstellar lines.
\citet{Harlaftis:1993a} suggest $\EBV=0.2$ based on 4428\,\AA\ and
6613\,\AA\ DIBs.  \citet{Shrader:1994a} also obtain $\EBV=0.2\pm0.1$
from the 5780\,\AA\ DIB.  Finally \citet{Callanan:1995a} measure
equivalent widths of $1.2\pm0.2$\,\AA\ for the Na\,D lines and
$0.5\pm0.1$\,\AA\ for the 5780/5788/5797\,\AA\ group.  Using the
relations of \citet{Barbon:1990a} and \citet{Herbig:1975a} they deduce
$\EBV=0.3\pm0.1$.  \citet{Chevalier:1995a} examine 12 stars in the
field surrounding \novaper\ and find that their color-color diagram
needs to be dereddened by $\EBV = 0.40\pm0.07$ in order to match the
unreddened main sequence.  This method assumes that all these stars,
and \novaper, are subject to the same reddening; it would be
appropriate if all are above the main plane of extinction.  Even if
some are not, but \novaper\ is (as seems likely given the estimated
distance for \novaper), then this value becomes a lower limit.
\citet{Shrader:1997a} fit absorbed power-laws to \ASCA\ spectra from
the 1993 August mini-outburst to obtain $N_{\rm
H}\simeq1.6\times10^{21}$\,cm$^{-2}$, corresponding to
$\ebv=0.28\pm0.08$.  Finally, and most recently, \citet{Gelino:2003a}
derive $\ebv=0.24\pm0.03$ from fitting the {\em quiescent} spectral
energy distribution, but this value is somewhat sensitive to
assumptions about the amount and color of disk light.  There is
clearly little close consensus on the reddening value, and even the
method which is usually most precise, fitting the 2175\,\AA\ feature,
yields equivocal results.  Consequently, based on the spread in
measured values we adopt $\EBV=0.35\pm0.10$.  This choice is rather
arbitrary, but is representative of the scatter between the derived
estimates.  It is completely consistent with the estimate of 0.36 from
dust maps.
\subsubsection{\novavel}
\citet{DellaValle:1997a} present several arguments for a low reddening
in \novavel, $\ebv\sim0.2$.  From the strength of Na\,D absorption
they derive an estimate of $\ebv=0.18$, with an upper limit of
$\ebv=0.23$.  Assuming that the intrinsic $(B-V)$ color is similar to
other LMXBs implies a consistent value of $\ebv\sim0.23$ (although see
caution above).  We follow these authors in adopting
$E(B-V)=0.20\pm0.05$.  The column density is not well constrained, but
probably $\la1.2\times10^{21}$\,cm$^{-2}$ \citep{Kubota:1998a}, which
implies $\ebv \la 0.27$, consistent with the optical estimates.
Finally dust maps predict $\ebv \la 0.19$, consistent with other
estimates.

\subsubsection{\novaophb}
The reddening of \novaophb\ has been estimated by
\citet{DellaValle:1994a} as $\ebv=0.9\pm0.2$ based on a number of
arguments giving roughly consistent values.  The strength of the Na\,D
lines implies $\ebv \ga 0.75$ (see the note on saturation of these
lines above). The strength of DIBs suggests $\ebv=0.6$--0.9.
Arguments based on the outburst $(B-V)$ color, and the Balmer
decrement give somewhat higher values, but are model dependent as
discussed above.  The interstellar hydrogen column density has been
estimated at $N_{\rm H}=4\times10^{21}$\,cm$^{-2}$ (Tanaka 1993),
implying $\ebv = 0.5$--$0.9$, which is consistent with the optical
estimates.  Finally, dust maps imply $\ebv \la 1.08$.  In summary, we
see no compelling reason to amend the estimate of
\citet{DellaValle:1994a}, leaving this as the largest and most
uncertain reddening in our sample.
\subsubsection{XTE\,J1859+226}
\citet{Hynes:2002a} estimated $\ebv=0.58\pm0.12$ using the well
defined 2175\,\AA\ interstellar feature.  \citet{Wagner:1999a}
obtained $\ebv=0.8\pm0.4$ from the 5780\,\AA\ DIB.  Estimates of the
hydrogen column density included $N_{\rm H} =
0.8\times10^{22}$\,cm$^{-1}$ (\citealt{dalFiume:1999a} from \SAX\
data) and $N_{\rm H} = 1.1\times10^{22}$\,cm$^{-2}$
(\citealt{Markwardt:2001a}; \RXTE).  This column density corresponds
to $\ebv=0.9$--$2.5$, allowing for the uncertainty and variance in the
gas-dust ratio. Even with this uncertainty, and that in the 2175\,\AA\
determination, there is then a discrepancy.  In contrast, dust maps
imply an even lower value, $\ebv \la 0.49$.  The 2175\,\AA\ value thus
represents a reasonable middle-ground, so we will use this estimate,
and the derived dereddened spectra from \citet{Hynes:2002a}.
%
%
\section{Distance Estimates}
\label{DistanceSection}
The determination of distances to BHXRTs is even more problematic than
their reddenings, since there are no direct measures of
distance.  Ideally, we would seek some kind of standard candle to
determine the distance.  For neutron star LMXBs, a standard candle
does exist in the form of radius expansion type-I X-ray bursts.  No
such analog exists for black holes, and any attempt to use the
outburst luminosity, as done by some authors, runs into a circular
logic problem again, that we would be artificially forcing uniformity
on our sample before comparing them.  The only reasonably robust
method that has been applied to our sample is to use the quiescent
companion star.  System parameters provide estimates of the
companion's temperature and radius, and hence absolute magnitude,
albeit with a large uncertainty in many cases.  Given the observed
magnitude, estimated extinction and the fraction of quiescent light
due to the disk (the veiling) we can then derive a distance.  The
errors are likely to be large, but are representative of the real
uncertainty in knowing the distance, and other methods are likely to
be even worse.  The best method to assess the errors is to perform a
simple simulation, constructing a population of model binaries
consistent to within the observational uncertainties and derive a
distance for each one, and hence a mean distance and standard
deviation.  This method has recently been applied to a number of our
sources using IR data.  This is preferable because the IR
derived distance is less sensitive to uncertainty in extinction or
companion temperature than the optical is.  The best estimate is for
\novamon\ for which \citet{Gelino:2001b} derive $1.16\pm0.11$\,kpc.
For \novamus\ \citet{Gelino:2001a} quote $\sim5.1$\,kpc with no
uncertainty, but this estimate was subsequently refined with
simulations as above to $5.89\pm0.26$\,kpc \citep{Gelino:2001c}.  For
\novaper\ \citet{Gelino:2003a} estimate $2.49\pm0.30$\,kpc and finally
for \novavel\ \citet{Gelino:2002a} derive $3.82\pm0.27$\,kpc.  Of
course, these are not the only estimates for these sources.  However
we believe they are the most thorough and reflect the most up to date
parameter estimates.  These estimates are broadly consistent with
previous values.  Some concerns do still exist for these objects in
that these estimates, and their quoted uncertainties, depend on the
limits placed upon the veiling.  The works cited generally assume
rather small veilings in the IR, and while this assumption is widely
made, its validity is questionable \citep{Hynes:2003a}.  The most
immediate effect of increasing the veiling is to reduce the apparent
brightness of the companion star, implying a larger distance.  It is
therefore quite possible that the distances are underestimated
somewhat.  For more detailed discussion of these issues see
\citet{Jonker:2004a}.

For the other two of our selected sources, the information necessary
to use the quiescent companion as a standard candle is simply not all
available yet, so in these cases we make use of other indicators from
the literature.

The distance to \novaophb\ is the poorest determined in our sample.
\citet{DellaValle:1994a} estimate 2\,kpc from relations between Na\,D
equivalent width and distance.  Since the Na\,D lines could well be
saturated for the relatively high extinction of this object, this
distance is really only a lower limit, even assuming that the mean
relation is valid along this line of sight.  Based on the peak optical
brightness, these authors estimate an upper limit of 2.8\,kpc, but as
noted above we should be cautious about using such measurements for
our purposes.  In this case, however, we have little other information
to go on and follow \citet{DellaValle:1994a} in adopting
$2.4\pm0.4$\,kpc for \novaophb.  The true uncertainty may well be
larger, however, and the distance could be significantly greater.

The distance to \novavulb\ is also problematic.  \citet{Zurita:2002a}
estimated $d=11$\,kpc from both the brightness of the outburst and
that of the quiescent companion.  The latter estimate was, however,
based largely upon guesses for a number of binary parameters including
the orbital period and companion spectral type.
\citet{Filippenko:2001a} report spectroscopy indicating an orbital
period longer than assumed by \citet{Zurita:2002a} and an earlier
spectral type.  Both effects will tend to make the quiescent companion
more luminous and hence push the source to a larger distance.  We have
done a simulation similar to those discussed above to test these
effects, and estimate that the quiescent brightness implies a distance
of $\sim21\pm5$\,kpc assuming the constraints of
\citet{Filippenko:2001a}.  This value depends on some uncertain or
totally unknown parameters, however, such as the inclination and mass
ratio.  The disk veiling is also ignored and a significant disk
contribution would push the distance even higher.  The faintness in
quiescence therefore suggests a very large distance, large enough to
conflict with other estimates.  Outburst X-ray observations place a
likely constraint of 5--13\,kpc, based on a combination of spectral
and timing information \citep{Markwardt:2001a}.  Fits to the
optical--UV spectral energy distribution suggested 4.6--8.0\,kpc
\citep{Hynes:2002a}, assuming a plausible range of possible disk radii
and inclinations.  In view of the extreme uncertainty about the system
parameters, in particular the companion star's effective temperature,
we believe these outburst estimates are currently more reliable, but
that we have no really trustworthy distance for this object.  We adopt
a working estimate of $8\pm3$\,kpc, which encompasses most of the
parameter space proposed thus far, but note that a larger distance
remains a possibility.
%
%
\section{Satellite Spectra}
\label{SatelliteSection}
\begin{deluxetable*}{llrllcl}
\tablecaption{Satellite spectra included in the sample.\label{SatelliteTable}}
\startdata
\hline
\noalign{\smallskip}
Object & Date & Outburst & Facility & Wavelength  & Plotted\tablenotemark{a} & Reference\tablenotemark{b} \\
       &      & Day      &          & range (\AA) &                          &           \\
\noalign{\smallskip}
\novamon & 1975 Sep 28--30  & 46--48 & ANS      & 1550--3300 & Y &  1 \\  
\noalign{\smallskip}
\novamus & 1991 Jan 17--21 &   2--6 & IUE      & 1150--3300 & A & 2 \\
         & 1991 Jan 24     &      9 & IUE      & 1150--3300 & A & 2 \\
         & 1991 Jan 29     &     14 & IUE      & 1150--3300 & N & 2 \\
         & 1991 Feb  6     &     22 & IUE      & 1150--3300 & A & 2 \\
         & 1991 Feb 12     &     28 & IUE      & 1150--3300 & A & 2 \\
         & 1991 Feb 18     &     34 & IUE      & 1150--3300 & A & 2 \\
         & 1991 Feb 25     &     41 & IUE      & 1150--3300 & A & 2 \\
         & 1991 Feb 28     &     44 & IUE      & 1150--3300 & A & 2 \\
         & 1991 Mar 15     &     59 & IUE      & 1150--3300 & N & 2 \\
         & 1991 Mar 27--28 & 71--72 & IUE      & 1150--3300 & N & 2 \\
         & 1991 Apr 22     &     97 & IUE      & 1150--3300 & N & 2 \\
         & 1991 May 15     &    120 & HST/FOS  & 1580--4810 & Y & 3 \\
\noalign{\smallskip}
\novaper & 1992 Aug 16     &      2 & IUE      & 1150--3300 & A & 4 \\
         & 1992 Aug 20     &      6 & IUE      & 1150--3300 & A & 4 \\
         & 1992 Aug 26     &     10 & IUE      & 1150--3300 & A & 4 \\
         & 1992 Sep  2     &     19 & IUE      & 1150--3300 & A & 4 \\
         & 1992 Sep 10     &     27 & IUE      & 1150--3300 & A & 4 \\
         & 1992 Sep 20     &     37 & IUE      & 1150--3300 & A & 4 \\
         & 1992 Sep 27     &     44 & IUE      & 1150--3300 & A & 4 \\
\noalign{\smallskip}
\novavulb & 1999 Oct 18    &      2 & HST/STIS & 1150-10\,200 & Y & 5 \\
\         & 1999 Oct 27    &     11 & HST/STIS & 1150-10\,200 & Y & 5 \\
          & 1999 Nov  6    &     21 & HST/STIS & 1150-10\,200 & Y & 5 \\
          & 2000 Feb  8    &    115 & HST/STIS & 1150-10\,200 & Y & 5 \\
          & 2000 Mar  5    &    141 & HST/STIS & 1150-10\,200 & Y & 5 \\
\noalign{\smallskip}
\hline
\enddata
\tablenotetext{a}{Spectra shown in Fig.~\ref{OptSEDFig} or Fig.~\ref{UVSEDFig}.  Y indicates a spectrum plotted individually, A means the spectrum is inclued in an average and N that the spectrum is omitted.}
\tablenotetext{b}{References: 1. Wu et al.\ (1976), 2. Shrader \&
  Gonzalez-Riestra (1993), 3. Cheng et al.\ (1992), 4. Shrader et al.\
  (1994), 5. Hynes et al.\ (2002)}
\end{deluxetable*}

\subsection{\ANS}
\label{ANSSection}
\novamon\ was observed by the Ultraviolet Experiment onboard the
Astronomical Netherlands Satellite (ANS) around day 47 after X-ray
peak \citep{Wu:1976a}; a second observation was made around day 225
\citep{Wu:1983a} which we do not use as it was obtained beyond the
main exponential decay phase of the outburst.
\citep{Wu:1976a} tabulate the data from the earlier observation.  From
these we constructed an average using only the six observations when
the broad 1550\,\AA\ bandpass was used to minimize the impact of the
possible C\,{\sc iv} emission line.
\subsection{\IUE}
\label{IUESection}
\IUE\ spectra were obtained for two of our sources: \novamus\
\citep{Shrader:1993a} and \novaper\ \citep{Shrader:1994a}.  Both
sources were observed with the low-resolution short- and
long-wavelength prime cameras (SWP and LWP).  We obtained all of these
data from the archive, using the {\sc newsips} reprocessed results
\citep{Nichols:1996a}.

For \novamus, good coverage was obtained throughout the outburst,
although the later data are of poorer quality.  We assembled three
representative average spectra.  For the early outburst, we combined
spectra from 1991 January 17--24; no obvious spectral differences are
visible between these.  We produced a second epoch spectrum from
February 6--28.  This corresponds to some way down the decline, but
before the secondary maximum in the X-ray lightcurve.  For the third
sample we use a single epoch, March 27--28, within the secondary
maximum, which was seen in the UV data \citep{Shrader:1993a}.  At
later epochs, the \IUE\ data are of poor quality, and the \HST\
spectrum (see next section) is of more use.  We did not attempt to
re-derive a reddening for \novamus\ from the \IUE\ data as the \HST\
spectrum of this source is of much higher quality.

For \novaper, the coverage was less complete; the first two months
were observed regularly, but there was only one more observation about
a year later.  We assembled two average early spectra from 1992 August
16--26 and 1992 September 2--27 respectively.  The last visit lies
beyond the main exponential decay phase of the outburst.
Since the reddening derived by \citet{Shrader:1994a} from the
2175\,\AA\ feature was inconsistent with other estimates, we combined
these two spectra and reanalyzed them to attempt to resolve the
discrepancy.  \citet{Shrader:1994a} derived $\ebv=0.40\pm0.06$, but
did so by assuming the entire \IUE\ spectrum was a power-law.
Consequently there is a danger that the derived reddening may be
biased by a failure of the model.  As for \novavulb\
\citep{Hynes:2002a}, we tried using a range of different wavelength
ranges and both a power-law and irradiated disk model in order to test
for such sensitivities.  We do indeed find that a wide range of
reddening values are derived, 0.29--0.48, dependent mainly on the
choice of fitting range.  As can be seen from the fit by
\citet{Shrader:1994a}, a power-law model does fail both at short and
long wavelengths.  We find that using the 1650--3120\,\AA\ range
adopted for \novavulb, significant systematic residuals are seen at
the longer wavelengths.  Fits with this region ($\lambda > 2500$\,\AA)
included produce large values of $\ebv = 0.37-0.48$, whereas if this
region is excluded a range of 0.29--0.38 is found.  The range above
2500\,\AA\ contains little if any information on the 2175\,\AA\
feature, and the discrepancy with other estimates is also largely
resolved if it is excluded, so we choose to use fits without this
region.

\subsection{\HST}
\label{HSTSection}
\HST\ spectra from the main part of the outburst are available for two
of our sources: \novamus\ \citep{Cheng:1992a} and \novavulb\
\citep{Hynes:2002a}.

The data on \novamus\ were obtained with the Faint Object Spectrograph
(FOS) fairly late in the outburst, but still in the
exponential decay phase.  We obtained these data from the
ST-ECF archive.  They were recalibrated by the archive using the {\sc
poa\_calfos} pipeline \citep{Alexov:2001a} and the final reference
files.  As for other sources, we reanalyzed the 2175\,\AA\ feature.
\citet{Cheng:1992a} found a good fit using a power-law model and we
find the same; the irradiated disk model which worked better for
\novavulb\ provides a very poor fit for \novamus.  A reddened
power-law fit to the 1574--3500\,\AA\ range (chosen to exclude any
weak Balmer jump contribution) gave $\ebv=0.30$, $\alpha = 0.52$.  The
reddening estimate is not significantly different to that of
\citet{Cheng:1992a}, but the derived spectrum is somewhat bluer.
This is due to the different extinction curve used; we use
that of \citet{Fitzpatrick:1999a}, whereas \citet{Cheng:1992a} used
the \citet{Seaton:1979a} curve.  The difference between these curves
is shown by \citet{Fitzpatrick:1999a}, and the different spectral
slopes derived is representative of the real uncertainty in the
process (see Section~\ref{DereddenVarSection} for more discussion of
this point.)  We adopt this value, with 20\,percent uncertainty, as a
reddening measure for \novamus: $\ebv=0.30\pm0.06$.  We also repeated
the fit with the full dataset, 1574--4816\,\AA, and with all the
wavelength ranges used for \novavulb.  In every case, we derive $\ebv$
consistent with the range $\ebv=0.30\pm0.06$.

A full analysis of data obtained with the Space Telescope Imaging
Spectrograph (STIS) of \novavulb\ along the lines here has already
been presented by \citet{Hynes:2002a}, we simply include selected
results of that work for comparison with the other sources presented
here.  These spectra span the full 1150--10\,300\,\AA\ range at
several epochs through the outburst.  These are the only data on this
source included.
%
%

\section{Ground-based Optical Spectra}
\label{OpticalSection}
\begin{deluxetable*}{llrllcl}
\tablecaption{Ground-based spectra included in the sample.
\label{GroundDataTable}}
\startdata
\hline
\noalign{\smallskip}
Object & Date & Outburst & Facility & Wavelength       & Plotted\tablenotemark{a} & Reference\tablenotemark{b} \\
       &      & day      &          & range used (\AA) &                          &           \\
\noalign{\smallskip}
\novamon & 1975 Sep 2--4  &   20--22 & Palomar, Hale 5\,m & 3200--10\,350 & Y & 1 \\  
         & 1975 Dec 8--11 & 117--120 & Palomar, Hale 5\,m & 3200--10\,350 & Y & 1 \\  
\noalign{\smallskip}
\novamus & 1991 Jan 15    &        0 & ESO, NTT    & 4350--7500 & Y & 2 \\
         & 1991 Jan 18    &        3 & ESO, 1.5\,m & 3800--8600 & A & 3 \\
         & 1991 Jan 19    &        4 & ESO, 1.5\,m & 3800--7400 & A & 3 \\
         & 1991 Jan 20    &        5 & ESO, 1.5\,m & 6300--8600 & A & 3 \\
         & 1991 Jan 23    &        8 & ESO, NTT    & 4200--6200 & A & 3 \\
         & 1991 Feb 15    &       31 & ESO, 1.5\,m & 3850--7450 & Y & 3 \\
         & 1991 Feb 21    &       37 & ESO, 1.5\,m & 3850--9850 & N & 3 \\
         & 1991 Mar 6     &       50 & ESO, NTT    & 4000--8150 & Y & 3 \\
         & 1991 Mar 21    &       65 & ESO, 2.2\,m & 4050--9050 & N & 3 \\
         & 1991 Mar 25    &       69 & ESO, NTT    & 6100--8100 & N & 3 \\
         & 1991 May 5     &      110 & ESO, 1.5\,m & 4000--7100 & N & 3 \\
         & 1991 May 19    &      124 & ESO, 3.6\,m & 4200--9850 & Y & 3 \\
\noalign{\smallskip}
\novaper & 1992 Aug 17    &        3 & Lowell Perkins, 1.8\,m & 3800--8500 & Y & 4 \\
         & 1992 Aug 18    &        4 & Lowell Perkins, 1.8\,m & 4850--6800 & N & 4 \\
         & 1992 Sep 25    &       42 & Lowell Perkins, 1.8\,m & 4850--6800 & N & 4 \\
         & 1992 Oct 10    &       57 & Lowell Perkins, 1.8\,m & 4850--6800 & Y & 4 \\
         & 1992 Nov  4    &       82 & Lowell Perkins, 1.8\,m & 4850--6800 & N & 4 \\
         & 1992 Dec 15    &      123 & Lowell Perkins, 1.8\,m & 4850--6800 & Y & 4 \\
         & 1993 Jan 21    &      160 & Lowell Perkins, 1.8\,m & 4850--6800 & N & 4 \\
         & 1993 Feb 11    &      181 & Lowell Perkins, 1.8\,m & 4850--6800 & Y & 4 \\
\noalign{\smallskip}
\novavel & 1993 Nov 18    &       64 & ESO, 2.2\,m & 3600--9050 & Y & 5 \\
\noalign{\smallskip}
\novaophb & 1993 Oct 5--8 &     5--8 & ESO, 1.5\,m \& 2.2\,m 2& 4600--6700 & Y & 6 \\
\noalign{\smallskip}
\hline
\enddata
\tablenotetext{a}{Spectra shown in Fig.~\ref{OptSEDFig}.  Y indicates a spectrum plotted individually, A means the spectrum is inclued in an average and N that the spectrum is omitted.}
\tablenotetext{b}{References: 1. \citet{Oke:1977a},
  2. \citet{DellaValle:1991a}, 3. \citet{DellaValle:1998a},
  4. \citet{Shrader:1994a}, 5. \citet{DellaValle:1997a}, 6. \citet{DellaValle:1994a}}
\end{deluxetable*}

\subsection{\novamon}
The highest quality calibrated spectra for \novamon\ were presented by
\citet{Oke:1977a} who show spectrophotometry from several epochs, of
which two, around 21 and 118 days after outburst peak respectively,
were obtained during the exponential decay phase of the
outburst.  The original data are no longer available, so these two
spectra were digitized from the paper and included in our sample.  The
first observation occurred close to the ANS UV observations.
\citet{Gull:1976a} present two low-resolution SEDs of the optical.
These are of poorer quality than those of \citet{Oke:1977a}, and are
only shown with an arbitrary flux scale, so we do not include these.
\subsection{\novamus}
A large sample of optical spectra were presented by
\citet{DellaValle:1991a} and \citet{DellaValle:1998a}.  These span
days 0--124 days after outburst peak.  The spectrum included in
\citet{DellaValle:1991a} was digitized; those from the later paper
were kindly supplied by the authors.  For the latter, the reddening
correction pre-applied was removed before dereddening them as
described above for consistency.  For further analysis, we merged the
spectra from 1991 Jan 18--23.  We also rejected the 1991 Mar 25
spectrum which has limited wavelength coverage, and a very different
slope to all others considered, even 1991 Mar 21.  Some portions at
the extreme near-IR and near-UV ends of some spectra show abrupt
and non-repeatable transitions and these were also excluded.
\subsection{\novaper}
\citet{Shrader:1994a} present spectra from 1992 August 16 to
September 25.  We include an absolutely fluxed spectrum from 1992
August 17, from this dataset, which was provided by the author.  The
remaining epochs were digitized from a figure also provided by the
author as the original data were not available.
\subsection{\novavel}
\novavel\ suffered both from a relatively short exponential phase of
outburst, $\sim150$\,days \citep{Bailyn:1995a}, and from a late
discovery.  The only spectrum from the exponential decay phase
was presented by \citet{DellaValle:1997a}, spanning 3500--9500\,\AA.
This spectrum was digitized, placed on an absolute flux scale
(assuming 1 unit $=$ $10^{-16}$\,erg\,cm$^{-2}$\,s$^{-1}$; Della
Valle, private communication), and corrected to our preferred
dereddening solution.
\subsection{\novaophb}
One flux calibrated spectrum from the exponential decay phase
was presented by \citet{DellaValle:1994a}.  This spectrum was
digitized and corrected to our preferred reddening value and
extinction curve as for other sources.
%
%
\section{The optical spectra}
\label{OpticalSEDSection}

\begin{figure*}
\includegraphics[scale=0.7]{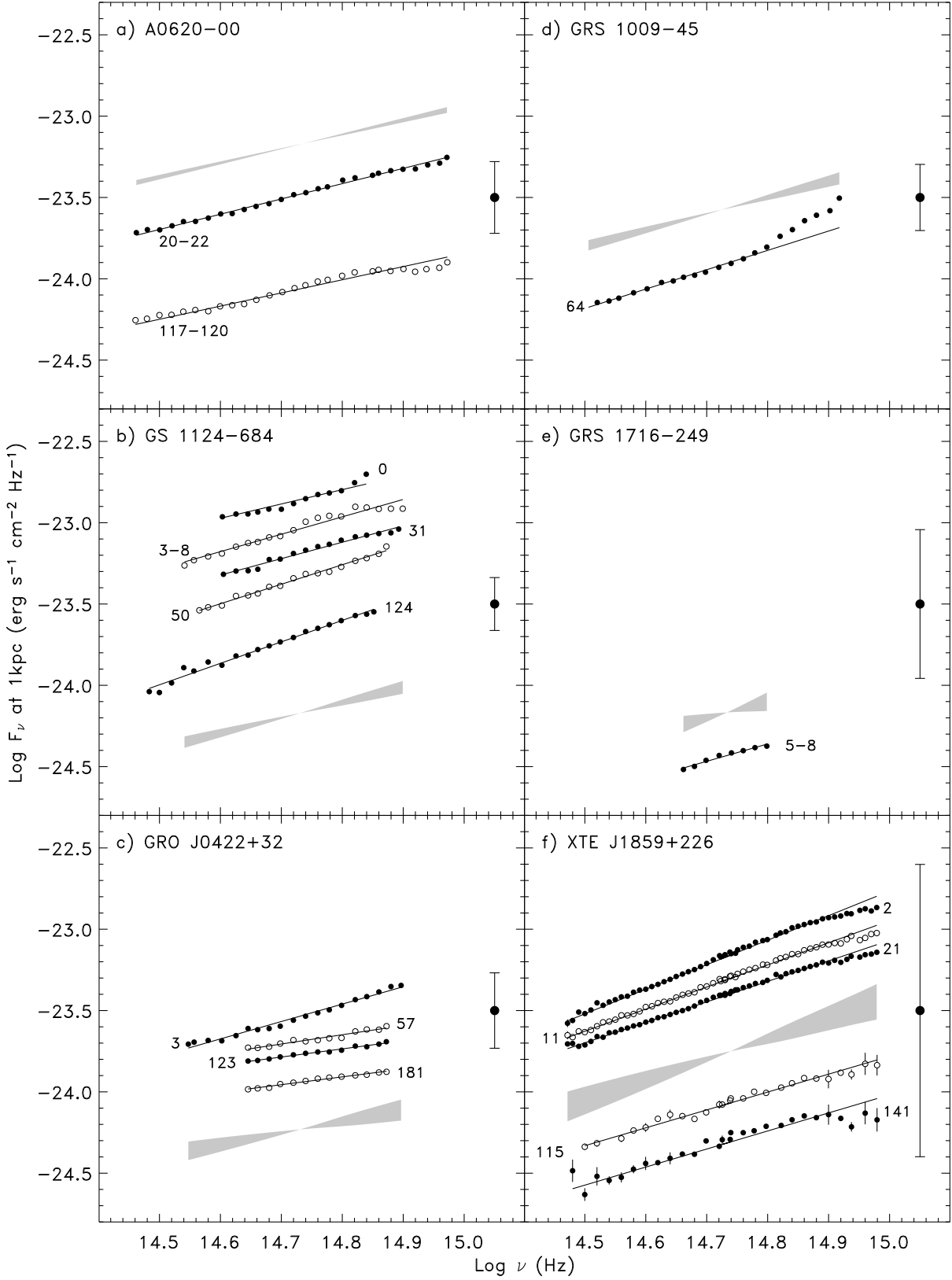}
\caption{Optical spectral energy distributions for all of our sources.
  Open and filled symbols are used to distinguish alternate epochs and
  have no other significance.  Numbers indicate the approximate number
  of days after X-ray outburst peak when each spectrum was obtained.
  The bow ties indicate the 1\,$\sigma$ uncertainty in spectral slope
  due to uncertain dereddening of the spectra.  The point with an
  error bar to the right of each panel indicates the uncertainty in
  absolute calibration for each object due to uncertain distance and
  absolute extinction.  Solid lines are power-law fits.}
\label{OptSEDFig}
\end{figure*}

Our combined sample of optical spectra are shown in
Fig.~\ref{OptSEDFig}.  For these purposes, satellite spectra have been
truncated at $\log \nu = 14.98$\,Hz, the approximate highest value
recorded for \novamon\ from the ground.  We have excluded some spectra
for which the flux calibration appears incorrect; when there are large
distortions which are not present at any other epoch; and others to
avoid duplication of two very similar spectra. 

It is clear that after applying the source selection criteria
described above, and subject to these additional restrictions, the
spectra define a relatively
homogeneous sample.  All are blue, with $F_\nu$ increasing toward the
UV.  All have an approximately power-law form.  Subject to the large
uncertainties in distance described above, all represent comparable
optical luminosities.

To perform a more quantitative comparison, we must parameterize and
fit the spectra.  At this stage we consider only very simple
characterizations; we will calculate model disk spectral energy
distributions in Section~\ref{ModelSection}, but UV coverage is
required to adequately constrain such models.  The two simple
characterizations widely used in the literature are a power-law or a
hot black body; the latter has an approximately power-law form on the
Rayleigh-Jeans tail, and so can also provide an adequate fit to the
data.  Given that a single temperature black body is likely a very
poor description of an accretion disk, the black body fit has limited
physical significance, but is useful for comparison with other work.
The temperatures that we derive are somewhat lower than other authors
have obtained in this way.  In the case of \novamon\ we estimate
13\,000--15\,000\,K, compared to 25\,000--30\,000\,K obtained by
\citet{Oke:1977a}. For both \novaophb\ and \novavel\ we estimate a
temperature of $\sim 17\,000$\,K, compared to published estimates of
20\,000\,K \citep{DellaValle:1997a} and 25\,000\,K
\citep{DellaValle:1993a} respectively.  The differences likely reflect
the difficulty of reliably constraining such high temperatures using
optical data alone, and in this regime large derived temperature
differences can result from quite small differences in the data and
its treatment (for example the wavelength range used in fitting.)  In
view of this difficulty, and the limited physical meaning of a single
temperature fit, we ignore the results of these fits from hereon;
suffice to say that {\em the spectra indicate a hot continuum,
sufficiently hot that optical observations have very limited
discriminatory power alone.}

\begin{deluxetable}{lcccc}[t]
\tablewidth{3in}
\tablecaption{Results of fits to the optical spectra.
\label{OptFitTable}}
\startdata
\hline
\noalign{\smallskip}
Object & Outburst & \multicolumn{2}{c}{Power-law\tablenotemark{a}} & Black body \\
       & day      & $\alpha$ & $\Delta\alpha$\tablenotemark{b}    & T (K) \\
\noalign{\smallskip}
\novamon &   20--22 & $0.94\pm0.01$ & $\pm0.07$ & 14\,700$\pm500$ \\  
         & 117--120 & $0.81\pm0.02$ & $\pm0.07$ & 12\,900$\pm500$ \\  
\noalign{\smallskip}
\novamus &        0 & $0.88\pm0.06$ & $\pm0.22$ & 14\,100$\pm800$ \\
         &     3--8 & $1.07\pm0.06$ & $\pm0.21$ & 17\,100$\pm500$ \\
         &       31 & $1.01\pm0.04$ & $\pm0.22$ & 17\,000$\pm400$ \\
         &       37 & $0.99\pm0.04$ & $\pm0.20$ & 15\,300$\pm600$ \\
         &       50 & $1.20\pm0.03$ & $\pm0.22$ & 19\,800$\pm800$ \\
         &       65 & $0.60\pm0.03$ & $\pm0.21$ & 11\,800$\pm300$ \\
         &      110 & $1.49\pm0.04$ & $\pm0.20$ & 31\,500$\pm3000$ \\
         &      124 & $1.32\pm0.04$ & $\pm0.20$ & 20\,100$\pm1000$ \\
\noalign{\smallskip}
\novaper &        4 & $1.07\pm0.03$ & $\pm0.36$ & 17\,700$\pm1000$ \\
         &        5 & $1.05\pm0.05$ & $\pm0.39$ & 18\,300$\pm800$  \\
         &       42 & $0.57\pm0.02$ & $\pm0.39$ & 13\,100$\pm300$  \\
         &       57 & $0.57\pm0.04$ & $\pm0.39$ & 13\,100$\pm500$  \\
         &       82 & $0.57\pm0.03$ & $\pm0.39$ & 13\,100$\pm400$  \\
         &      123 & $0.50\pm0.02$ & $\pm0.39$ & 12\,600$\pm300$  \\
         &      160 & $0.39\pm0.03$ & $\pm0.39$ & 11\,800$\pm300$  \\
         &      181 & $0.48\pm0.02$ & $\pm0.39$ & 12\,400$\pm200$  \\
\noalign{\smallskip}
\novavel &       64 &  $1.19\pm0.03$ & $\pm0.17$ & 17\,000$\pm700$  \\
\noalign{\smallskip}
\novaophb &       5 & $1.07\pm0.07$ & $\pm0.78$ & 17\,200$\pm900$  \\
\noalign{\smallskip}
\novavulb &       2 & $1.51\pm0.01$ & $\pm0.40$ & 28\,000$\pm600$ \\
          &      11 & $1.36\pm0.01$ & $\pm0.40$ & 22\,500$\pm400$  \\
          &      21 & $1.26\pm0.01$ & $\pm0.40$ & 19\,900$\pm400$  \\
          &     115 & $1.10\pm0.03$ & $\pm0.42$ & 17\,400$\pm600$  \\
          &     141 & $1.11\pm0.05$ & $\pm0.42$ & 17\,100$\pm900$  \\
\noalign{\smallskip}
\hline
\enddata
\tablenotetext{a}{Defined by $F_{\nu} \propto \nu^{\alpha}$.}
\tablenotetext{b}{Uncertainty due to uncertain $\ebv$.}
\end{deluxetable}

\begin{figure}
\includegraphics[scale=0.5]{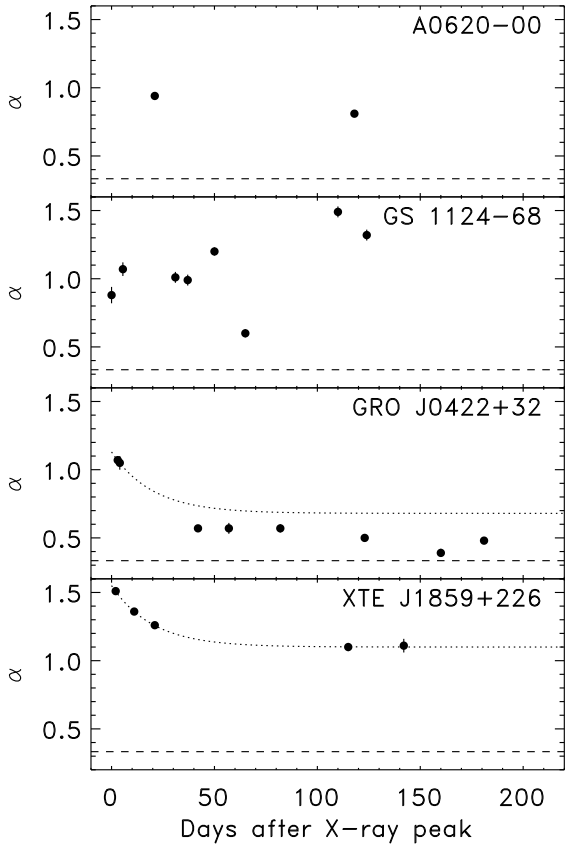}
\caption{Evolution of the optical power-law index ($F_{\nu} \propto
  \nu^{\alpha}$).  The dashed line corresponds to the canonical
  $\nu^{1/3}$ value for a steady-state viscously heated disk; all
  spectra appear steeper than this.  The dotted line is shows an
  exponential decay fitting the data for \novavulb, and is intended
  only to guide the eye in comparing these data with those for
  \novaper, for which the line has been offset downward by 0.45.}
\label{PLEvolutionFig}
\end{figure}

Power-law fits have previously been performed for some of our sources.
They provide a more empirical description of the data, but one which
more directly relates to its information content.  For \novamus\
\citet{King:1996a} found an initially blue spectrum, $\alpha\sim0.7$
(where $F_{\nu} \propto \nu^{\alpha}$), which softened to
$\alpha\sim0.2$ after the secondary maximum.  \citet{Cheng:1992a} also
remarked that this object exhibited a period at the beginning of the
outburst when the source had a steeper (bluer) spectrum than
later observed.  For \novaper, \citet{Shrader:1994a} find the optical
spectral index softened from 0.6 to 0.0 over the first three months of
the outburst, then stabilized.  \citet{King:1996a} found spectral
indices of between -0.1 and -0.4 for most of the outburst, although
the evolution is less clear in their data and they assumed a
relatively low extinction, $A_{V}=0.72$.  We fit power-laws to our
optical spectra and tabulate power-law indices in
Table~\ref{OptFitTable}.  These values differ somewhat from those
quoted above, as absolute values are rather sensitive to the assumed
$\ebv$.  Some spectral indices have also been plotted as a function of
time in Fig.~\ref{PLEvolutionFig}.  All of the values that we derive
fall within the approximate range $+0.5 < \alpha < +1.5$, and it
should be noted that {\em all} are steeper (bluer) than the canonical
$\alpha = 1/3$ of a steady-state viscously heated disk which is
sometimes assumed.  Some variations within this range may be due to
incorrect dereddening, but the full range cannot be explained in this
way, and the variations between repeated observations of a single
source indicate that there are real differences.  There is no single
`correct' value, and so the utility of assuming a standard spectral
slope (or photometric color) in estimating reddening values is
limited.  The observed variation corresponds to an intrinsic $(B-V)$
color of 0.00 to -0.24.  This range is small enough that it does have
some value for an approximate reddening estimate for sources with
relatively large reddening, but one should still be cautious, as we
have only obtained this degree of uniformity by {\it a posteriori}\
selection of sources by outburst behavior and orbital period.  For
example, the unusual source XTE~J1118+480 has a somewhat flatter
optical spectrum \citep{Hynes:2000a} and the long-period system
GRO~J1655--40 is much redder in the optical \citep{Hynes:1998a}

For \novamon\ there are too few points to draw any convincing
conclusions about evolution of the optical spectrum, other than that
little change appears to have occurred between the two epochs.  This
conclusion is borne out on longer timescales by the photometry of
\citet{Lloyd:1977a} who find approximately constant $B-V$ values
around $+0.25$ throughout the outburst, corresponding to a dereddened
color of $-0.1$, comparable to our spectral shapes.  For \novamus\
there are more data, but these exhibit rather large scatter.  The
spectra may be hardening, but it is hard to say with confidence, and
the photometric analyses of \citet{King:1996a} and
\citet{DellaValle:1997a} suggest the opposite behavior.  \novaper\ and
\novavulb\ show more coherent trends, but this may reflect better
sampling and/or data quality rather than an intrinsic difference.  The
latter in particular, being based solely on satellite data, should be
robust.  Both sources clearly become softer optically during the
outburst.  There appears to be rapid evolution during the first month,
then slower softening (at least for \novaper) later.  This evolution
has already been noted by \citet{Shrader:1994a}, although all of their
power-law indices are softer than ours due to a lower assumed
reddening.  The photometry of \citet{King:1996a} may also indicate an
initial rapid softening.  This behavior suggests that the optically
bright regions are cooling during the outburst.  We will defer more
detailed discussion until we consider the combined optical and UV
spectra of these sources (Section~\ref{ModelSection}).
Although the evolution is very similar for \novaper\ and \novavulb,
the actual values are rather different.  This difference may be real,
or may reflect an incorrect dereddening of the spectra.  The dotted
lines plotted in Fig.~\ref{PLEvolutionFig} are offset by 0.45.  At
late times \novaper\ is about a further 0.25 below the plotted line.
Uncertainty in reddening introduces an uncertainty in the spectral
slope of $\sim0.4$ for both sources, however, so most of the
difference in the optical slopes could be due to an incorrect $\ebv$.
We will reexamine the difference in the light of the UV spectra and
models in Section~\ref{ModelSection}.

We have invested some time in comparing the optical results in
isolation as these are often all that is available.  In addition these
results can be compared directly with optical photometry which
typically has better temporal sampling.  Based on optical results
alone, the objects appear relatively homogeneous and we cannot be
confident that apparent differences are not due to incorrect
dereddening.  Optical colors do give some useful information about
spectral {\em evolution} independent of reddening uncertainty.  They
do not allow us to distinguish between different models robustly
however; for that we must extend the data into the UV.  Based on the
analysis so far, \novaper\ and \novavulb\ exhibit definite and similar
softening during the outburst, whereas \novamon\ and \novamus\ exhibit
no clear trends, although very little data are available for the
former, and the latter exhibit large scatter.
%
%
\section{Optical and UV spectra}
\label{UVSEDSection}
\begin{figure*}
\includegraphics[scale=0.7]{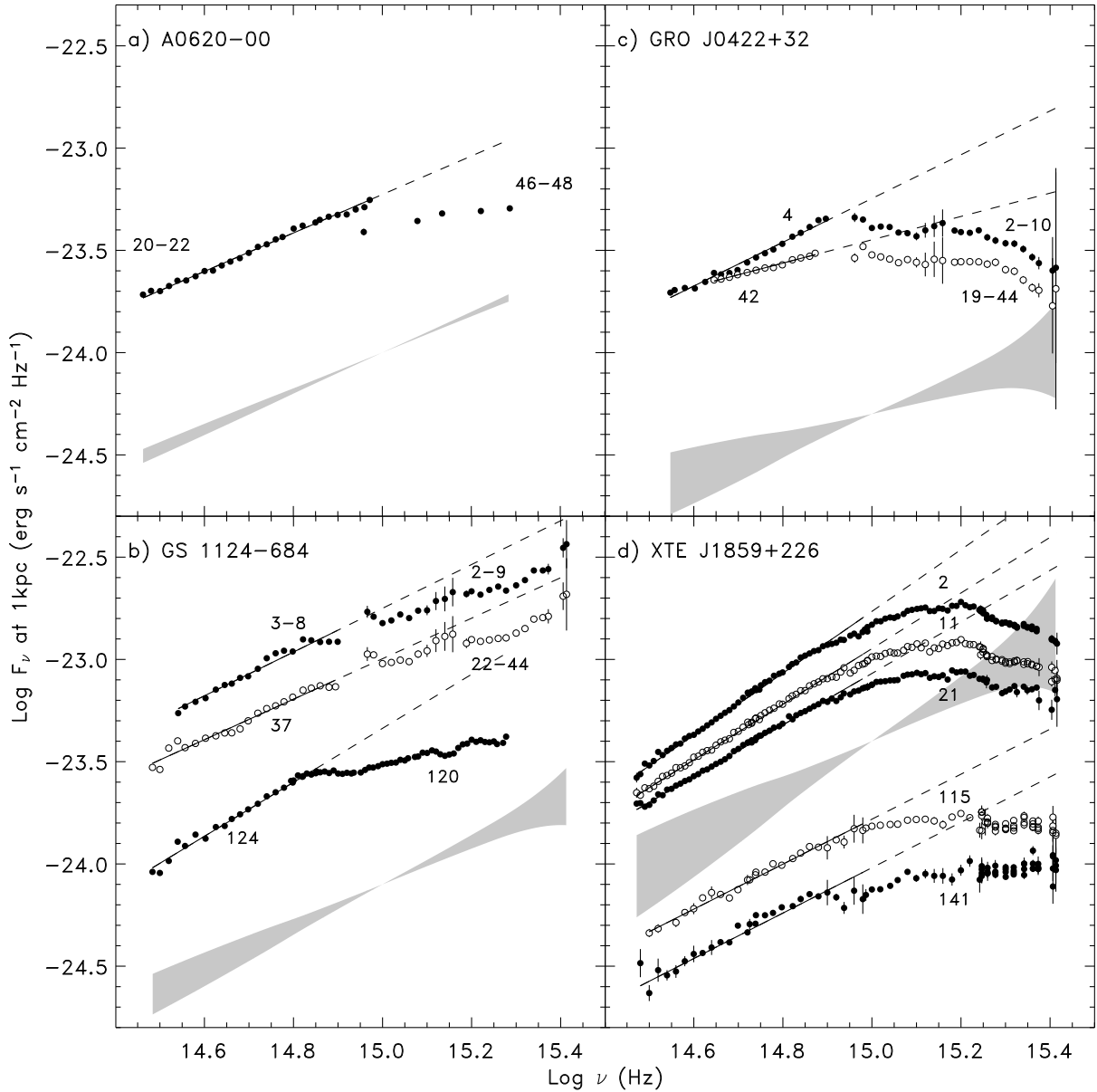}
\caption{Broad-band spectral energy distributions for the sources
  which have UV data.  Notation is as for Fig.~\ref{OptSEDFig}.  The
  distorted bow-ties indicate the uncertainty in the spectral shape
  introduced by uncertainty in $\ebv$.  These are calculated assuming
  a Galactic average extinction curve, and do not account for
  additional uncertainties introduced by variations in the extinction
  curve; the latter are illustrated in Fig.~\ref{UVExtCurveFig}.
  Dashed lines indicate extrapolations of power-law fits to the
  optical data.}
\label{UVSEDFig}
\end{figure*}

Our combined sample of broad-band spectra is shown in
Fig.~\ref{UVSEDFig}.  We have selected data for which both optical and
UV coverage is available.  Where possible we have used contemporaneous
data, but in some cases, particularly for \novamon, the time
difference is significant.

It can be seen that the power-law fits to the optical data do not
adequately extrapolate for any object.  All UV spectra flatten with
respect to the optical.  The degree of flattening varies, with
\novamon\ and \novamus\ continuing to rise more slowly in the UV,
while \novaper\ and \novavulb\ actually turn over and begin to fall.
One might question whether this difference is real or reflects
incorrect dereddening, but since we see both forms in \novavulb, it is
likely a real variation within the sample.  We can characterize this
distinction by dividing the spectra into two optical/UV states.  We
refer to spectra in which the far-UV spectrum continues to rise as
UV-hard, and ones in which it turns over and declines as UV-soft.

The presence of such a turnover is natural and is a consequence of the
cutoff in the disk temperature distribution at the disk edge (or at a
cooling front).  We will discuss the interpretation of these spectra
in Section~\ref{ModelSection}.  First, however, we examine more
carefully how the spectra are affected by uncertain dereddening.
%
%
\section{The impact of uncertain dereddening}
\label{DereddenVarSection}
\subsection{Optical}
The effect of uncertain dereddening is easiest to quantify in the
optical, as optical extinction curves are less diverse than those in
the UV, and the variation of extinction with wavelength is 
approximately monotonic.  Uncertainty in $\ebv$ will introduce an
uncertainty in the optical spectral slope, whereas an uncertain $\Av$
will affect the implied luminosity level.  Note that these effects are
semi-independent since $\Rv=\Av/\ebv$ may not always have the Galactic
average value of 3.1.  To quantify the effect on the spectral slope,
for each source we have shown in Fig.~\ref{OptSEDFig} a bow-tie
indicating the power-law slopes derived if $\ebv$ is allowed to vary
within the uncertainties estimated.  We have also included a
corresponding estimate of the change in power-law index,
$\Delta\alpha$, in Table~\ref{OptFitTable}.
\subsection{UV}
The effect on the UV spectrum is more complex, as the UV extinction
curve exhibits several features, and the relative strength of these
can vary.  The effect of uncertain $\ebv$, assuming a fixed extinction
curve, can be quantified in the same way as for the optical data,
although the form is not simply a variation in power-law slope.  This
effect is shown by the curved bow-ties in Fig.~\ref{UVSEDFig}. Note
that there is no increase in the uncertainty around the 2175\,\AA\
bump as this region is explicitly flattened out in dereddening the
spectra; a larger $\ebv$ would only be consistent with the data if the
bump were unusually weak and vice versa.

\begin{figure}
\includegraphics[angle=90,scale=0.35]{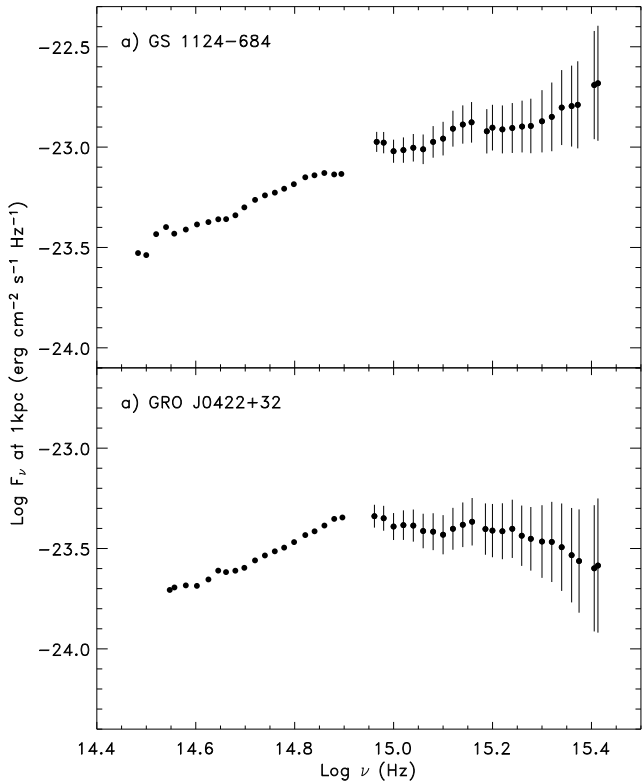}
\caption{Effect of uncertainty in the {\em shape} of the extinction
  curve on the dereddened spectra of \novamus\ and \novaper.  A
  similar plot for \novavulb\ was presented by \citet{Hynes:2002a}.
  Error bars on UV points indicate the uncertainty induced solely by
  extinction curve variance, following the recipe of
  \citet{Fitzpatrick:1999a}. }
\label{UVExtCurveFig}
\end{figure}

\citet{Fitzpatrick:1999a} has presented a convenient parameterization
of the variations in the extinction curve, and also presents a recipe
for quantifying the effect of the resultant uncertainty in the derived
SED.  \citet{Hynes:2002a} have already applied this analysis for
\novavulb.  We show comparable results for \novamus\ and \novaper\ in
Fig.~\ref{UVExtCurveFig}.  Note the uncertainty in the extinction
curve is less of a problem for \novamon\ for which a line-of-sight
curve was determined in parallel with the reddening determination
\citep{Wu:1983a}.
%
%
\section{Disk Modeling}
\label{ModelSection}
\subsection{The basic model}
We have shown that while a single power-law is an acceptable
description of most optical spectra, it does not adequately describe
the UV break, which seems to be ubiquitous.  This break is to be
expected.  We believe that the spectra included in this sample should
reflect the spectrum of an accretion disk in a hot state.  The hot
region should terminate at effective temperatures between 6\,000\,K
and 10\,000\,K, as regions at this temperature should be unstable.  If
such a cut-off occurs, then the optical spectrum will lie on the
Rayleigh-Jeans tail for most of the disk area hence exhibits a steeper
(bluer) spectrum than at higher frequencies, as observed.

We can attempt to quantify the disk temperature distribution using the
commonly used model for a disk heated by both viscosity and
irradiation and emitting locally as a black-body.  This model, its
origins, and original references, are summarized by
\citet{Hynes:2002a}, and we will not repeat the discussion.  An
alternative discussion is provided by \citet{Chakrabarty:1998a}.  We
use that model exactly as described by \citet{Hynes:2002a}, and with
the same definitions and terminology.  We emphasize that the model
actually only describes a disk with two forms of heating, one of which
drops off as $r^{-3}$ (viscous heating) and one as
approximately $r^{-2}$ (irradiation heating).  The exact
model used (based on \citealt{Vrtilek:1990a}) assumes a slightly
slower drop-off of the latter and is adopted for consistency with
earlier work and as a simple empirical description, rather than as a
theoretically motivated formulation (see \citealt{Dubus:1999a}).  In
practice the exact prescription adopted has little effect on the
optical and UV spectra, and so any error introduced in this way will
be less than that due to uncertainties in dereddening the spectra.
The essential discriminating evidence between these two effects is the
shape of the far-UV spectrum; a UV-hard spectrum which continues to
rise favors $r^{-3}$ heating, whereas a UV-soft spectrum with a clear
peak indicates $r^{-2}$ heating.  Beyond the peak, the UV-soft
spectrum will tend towards a $\nu^{-1}$ spectrum or steeper, but this
form will not be realized within the observable spectral range.  The
$r^{-3}$ component, which gives rise to the classical $\nu^{1/3}$
accretion disk spectrum \citep{Shakura:1973a}, is usually assumed to
be due to viscous heating, but could equally describe irradiation
which drops off in a similar way with radius.  Indeed this functional
form is commonly adopted in modeling irradiated disks in protostars
(e.g.\ \citealt{Hartmann:1998a}).  We will return to this alternative
interpretation in Section~\ref{CoronaSection}.

The analysis we will perform is virtually identical to that done by
\citet{Hynes:2002a} for \novavulb\ and we will directly compare to
those results without reanalyzing those data.  One important
difference, however, is that \novavulb\ is the only source for which a
full UV--optical spectrum was obtained quasi-simultaneously with \HST.
For the other objects in our sample having UV coverage, the optical
spectra were taken from the ground days or even weeks apart.  The
non-simultaneity compromises the reliability of the SEDs obtained, and
we will discuss the severity of this problem on a case-by-case basis.
In general it requires that we introduce an additional free parameter
representing the difference in flux levels between the two epochs, but
it should be remembered that spectral changes might have occurred as
well.

It is worth emphasizing that we explicitly assume there are no other
sources of flux besides the accretion disk; emission from a jet or
heating of the companion star are neglected.  For the sources and
wavelength region adopted, this is a reasonable assumption, and was a
major criteria in their selection.

\begin{deluxetable*}{llrcrclc}
\tablecaption{Parameters of disk fits to spectral energy distributions.
\label{ModelFitTable}}
\startdata
\hline
\noalign{\smallskip}
Object & Dates & \multicolumn{2}{c}{$T_{\rm irr,out}$ (K)} & \multicolumn{2}{c}{$T_{\rm visc,out}$ (K)} & \multicolumn{2}{c}{Normalization} \\
\noalign{\smallskip}
\novamon & 20--48 & 0 & 0--6\,000 & 7\,400 & 7\,000--7\,800 & 11.35 & 9.65--13.35 \\
\noalign{\smallskip}
\novamus & 2--9    & 0 & 0--6\,000 & 10\,000 & 8\,600--11\,400 & 0.85 & 0.61--1.26 \\
\novamus & 22--44  & 0 & 0--5\,000 & 7\,800  & 7\,000--8\,600 &  0.98 & 0.75--1.31 \\
\novamus & 120-144 & 7\,000 & 5\,000--9\,000 & 8\,400 & 8\,000--8\,800 & 0.23 & 0.19--0.27 \\
\noalign{\smallskip}
\novaper & 2--10  & 12\,600 & 11\,400--14\,200 & 5\,200 & 3\,000--5\,600 & 1.31 & 0.98--1.72 \\
\novaper & 19--44 &  9\,400 & 8\,400--11\,000 & 5\,000 & 4\,600--5\,600 & 1.98 & 1.28--2.78 \\
\noalign{\smallskip}
\hline
\enddata
\end{deluxetable*}

\subsection{\novamon}
\novamon\ has a high quality optical SED, and a well determined UV
extinction, so in principle it should be one of the best suited
objects to our study.  It suffers, however, from very crude wavelength
coverage in the UV and, most severely, from the lack of simultaneous
UV and optical observations.  We will compare optical data taken
around day 21 of the outburst with UV data from around day 47.  The 26
day interval between them may mean that the spectrum has changed
significantly.  A particular concern is that the UV observation
happened around the time of the onset of a secondary maximum in the
X-ray lightcurve (e.g.\ \citealt{Kuulkers:1998a}).  The optical
spectrum taken later, around day 118, is very similar to the early one
however, suggesting that little evolution actually occurred in the
optical spectrum, and that the comparison is valid.

\begin{figure}
\includegraphics[angle=90,scale=0.37]{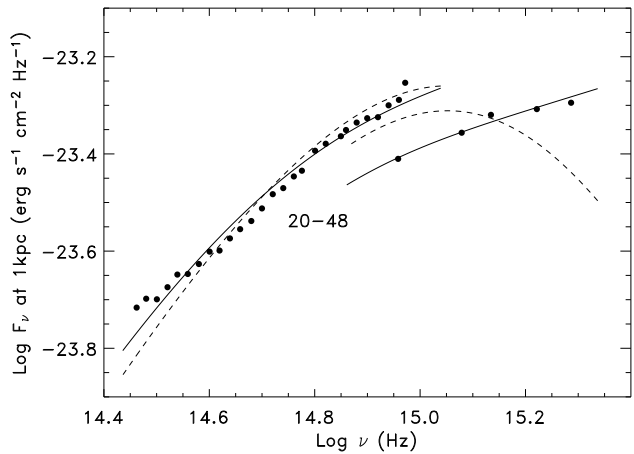}
\caption{Disk model fits to the spectra of \novamon.  Since the
  optical and UV data are non-simultaneous, the separation has been
  retained and models have been offset to fit.  The solid line shows
  the result of an unconstrained fit, which favors a disk heated only
  by viscosity.  The dashed line shows a fit constrained to $T_{\rm
  visc}=0$, i.e.\ a purely irradiated disk.  The latter clearly cannot
  fit well.  To improve the UV fit would require a large increase in
  the irradiation temperature and would then steepen the optical
  spectrum to an unacceptable degree.}
\label{NMonFitFig}
\end{figure}

Keeping these concerns in mind, we proceed to fit a disk model to the
SED, allowing for an arbitrary offset between the optical and UV data.
This approach is equivalent to assuming that the only difference
between the two epochs was in the flux level of the spectrum, rather
than the shape.  The fits are shown in Fig.~\ref{NMonFitFig}.  The
best fit model has no irradiative heating and corresponds to the
classic viscously heated steady-state disk.  For comparison we also
show a purely irradiated model.  This is an extremely poor fit, even
with an arbitrary offset between the optical and UV.  Improving the UV
fit would require an increase in the irradiation temperature and would
steepen the optical spectrum.  Of course some optical spectral
evolution may have occurred.  The later spectrum is flatter, however,
and other sources also tend to flatten as the outburst
proceeds. Flattening of the optical spectrum would further degrade the
irradiated fit, so it seems unlikely that the irradiated model can be
accommodated in this way.

The implied parameters (for the UV epoch) are sensible.  The outer
edge viscous temperature, $7400\pm400$\,K, is comparable to that found
in \novavulb\ late in outburst, and is close to the expected minimum
temperature of the hot phase of the disk \citep{Lasota:2001a}.  The
normalization can be interpreted as described by \citet{Hynes:2002a}.
Assuming a distance $d=1.16\pm0.11$\,kpc, and inclination
$i=40.75\pm3.0^{\circ}$ \citep{Gelino:2001b}, we estimate a disk
radius $r_{\rm out}=(1.8\pm0.3)\times10^{11}$\,cm, comparable to the
expected tidal truncation radius of the disk ($1.7\times10^{11}$\,cm).
The implied disk radius would increase to $2.0\times10^{11}$\,cm when
the model is normalized to fit the optical data.

We can also estimate the mass flow rate required to provide this
optical luminosity as described by \citet{Hynes:2002a}.  At the UV
epoch, we estimate
$\dot{M}=(0.9\pm0.3)\times10^{-7}$\,M$_{\odot}$\,yr$^{-1}$.  Assuming
an 11.0\,M$_{\odot}$ black hole \citep{Gelino:2001b}, and 10\,\%\
accretion efficiency, the Eddington limit corresponds to
$\dot{M}=2.5\times10^{-7}$\,M$_{\odot}$\,yr$^{-1}$, so we obtain a
mass flow rate a few tenths of the Eddington limit.

Given the uncertainties, the observations are thus consistent with a
lobe-filling, viscously heated disk.  Contrary to common assertions,
there seems to be no need for irradiative heating to provide
the energy radiated in the optical-UV.  This is not to say that
irradiation plays no role; it may still be crucial in stabilizing the
disk against a transition to a cool state.  The point is that the
main source of energy emitted is consistent with viscous heating
rather than irradiation.

Finally, we note that the long wavelength spectrum is flatter than the
model predicts.  This difference is relatively subtle, and may not be
of great significance.  However, a similar long-wavelength flattening
is seen in \novavulb\ \citep{Hynes:2002a} and may indicate a weak
contribution from synchrotron emission \citep{Brocksopp:2002a}.

\subsection{\novamus}
\novamus\ has better temporal coverage than \novamon, and generally
optical observations can be selected from within the time-period
spanned by multiple \IUE\ observations.  In addition, the \HST\
spectrum obtained later in the outburst is of higher quality, albeit
lacking the shortest wavelengths.  Two concerns are manifest here,
however.  Firstly, we have already seen that the power-law slopes
derived from the optical spectra scatter significantly.  The most
likely interpretation of this scatter would be that the flux
calibration is not totally reliable.  If this is correct then the
spectral fits are somewhat compromised.  Secondly there is a
sharp upturn in the far-UV, rather similar to that seen in \novavulb\
when the latter was corrected with the pure \citet{Fitzpatrick:1999a}
extinction curve.  \citet{Hynes:2002a} interpreted the upturn in that
case as due to a far-UV rise component of extinction which is weaker
than in the \citet{Fitzpatrick:1999a} extinction curve.  The same
could be true for \novamus, and similarly introduces uncertainty in
the dereddened far-UV spectrum.  We therefore omitted these points
from the fits.

\begin{figure}
\includegraphics[angle=90,scale=0.37]{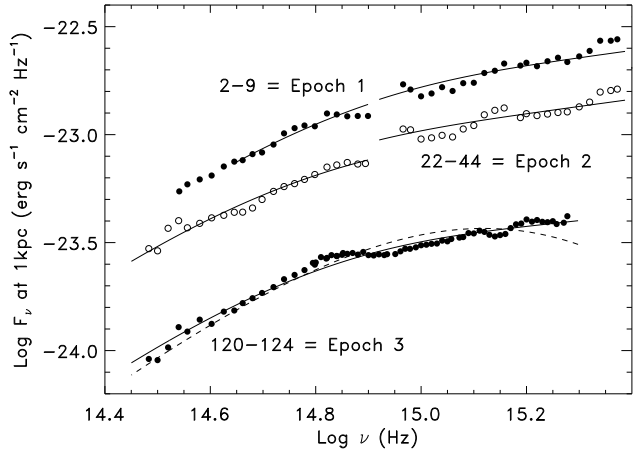}
\caption{Disk model fits to the spectra of \novamus.  For the first
  two epochs optical and UV data do not overlap or line up well so
  they have been fitted separately.  The last epoch spans a shorter
  period, and the data overlap, so no offset was included in this
  case.  The solid line shows the result of an unconstrained fit,
  which favors a disk heated only by viscosity.  The dashed line for
  Epoch 3 shows a fit constrained to $T_{\rm visc}=0$, i.e.\ a purely
  irradiated disk.}
\label{NMusFitFig}
\end{figure}

Disk model fits are shown in Fig.~\ref{NMusFitFig}, performed as for
\novamon.  \novamus\ is also best fitted by a model dominated by
viscous heating, exhibiting UV-hard spectra at all epochs.  A model
with purely irradiative heating would not fit.  The disk outer edge
temperature is comparable to that in \novamon\ at late times, but
higher earlier, dropping from $10\,000\pm2000$\,K to $\sim8\,000$\,K;
a similar decline was found by \citet{Cheng:1992a}.  Assuming
$d=5.89\pm0.26$\,kpc and $i=54^{+4^{\circ}}_{-1.5}$
(\citealt{Gelino:2001a}; \citealt{Gelino:2001c}) we obtain disk radii
of 1.4--3.0$\times10^{11}$\,cm, comparable to the expected tidal
truncation radius of $1.7\times10^{11}$\,cm.  The scatter inferred in
the radius may arise from the uncertainty in the optical spectral
slope, since this uncertainty will affect the derived temperature,
which in turn will be inversely correlated with the normalization.
These data thus also appear approximately consistent with a lobe
filling disk.  The higher temperatures early on may indicate that the
disk is indeed being tidally truncated, since the hot-phase of the
disk should be able to extend to below 10\,000\,K if it were
unbounded.

The temperatures we estimate are higher than for \novamon, even though
the disk radius is expected to be comparable.  A higher temperature
implies a larger mass flow rate (if the spectrum is indeed due to
viscous heating).  Note that this conclusion is largely independent of
distance uncertainties, provided both disks are tidally truncated.
\citet{Esin:2000a} indirectly provide support for this conclusion.
They compare the X-ray lightcurves and estimate a flux scaling factor
of $\eta\sim8$ for the ratio of \novamon\ to \novamus.  They argue
that the overall flux level of the two outbursts should be the same,
assuming that the Eddington scaled luminosity of the intermediate
state should be constant.  This is only expected, however, for
extreme parameters, for example a black hole mass in \novamon\ of just
4.5\,M$_{\odot}$ is assumed.  If we instead adopt current best
estimates of the parameters \citep{Gelino:2001a,Gelino:2001b}, then we
instead predict a flux scaling of $\eta=56\pm18$, if the Eddington
scaled luminosities were the same.  This scaling is uncertain, but
significantly larger than observed, implying that \novamon\ is
underluminous relative to \novamus, and hence that the accretion rate
in \novamus\ was likely larger.  \citet{Esin:2000a} also note that the
ratio of optical to X-ray flux is a factor of 1.6 larger in \novamon\
than in \novamus, and attribute this difference to more efficient
irradiation of the outer disk in \novamon.  There is another
explanation, however.  Both the optical and X-ray bandpasses lie at
the extreme ends of the spectral energy distribution, on the
Rayleigh-Jeans and Wien tails respectively.  If the mass flow rate is
higher, the effect is to increase the temperature throughout the disk
and shift the SED out of the optical bandpass and into the X-ray one.
Such a difference is supported by the hotter optical/UV spectra
inferred in \novamus.  This spectral shift will naturally decrease the
optical to X-ray flux ratio without invoking any other changes.  For
example, consider two viscously heated disks around 10\,M$_{\odot}$
black holes, with inner and outer radii of 3\,R$_{\rm sch}$ and
$2\times10^{11}$\,cm respectively.  We adjust the mass flow rate to
give outer disk temperatures of 8\,000\,K and 12\,000\,K respectively
and calculate the ratio of $V$ band to X-ray flux (1.3--7.5\,keV).
The ratio can change by at least a factor of two and possibly more,
depending on the relative weighting across the X-ray bandpass and the
effect of spectral hardening on the X-ray spectrum.  This effect is
more than adequate to explain the difference in the flux ratios
between \novamon\ and \novamus\ without any irradiation effect.

If we assume a 6.95\,M$_{\odot}$ black hole in \novamus, then we
estimate mass flow rates of
$(2.0\pm1.2)\times10^{-6}$\,M$_{\odot}$\,yr$^{-1}$,
$(8.5\pm1.2)\times10^{-7}$\,M$_{\odot}$\,yr$^{-1}$, and
$(1.3\pm1.2)\times10^{-7}$\,M$_{\odot}$\,yr$^{-1}$ for epochs 1--3
respectively (comparable to that obtained by \citet{Misra:1999a}
assuming similar parameters).  These values are superficially somewhat
problematic, as the Eddington limit should be reached for
$\dot{M}=1.6\times10^{-7}$\,M$_{\odot}$\,yr$^{-1}$, assuming 10\,\%\
accretion efficiency.  Even with higher efficiencies, the early phase
of the outburst should thus have involved super-Eddington accretion.
This is not a fatal problem, however, since the Eddington limit
strictly only applies to spherical accretion and a geometrically thin
disk could deliver mildly super-Eddington accretion rates.  Such a
super-Eddington model was indeed considered for \novamus\ by
\cite{Misra:1999a}, and \citet{Esin:2000a} also assumed a peak
accretion rate somewhat above the Eddington limit.  A more pertinent
question is whether such a large accretion rate can be reconciled with
the observed thermal (disk) component in the X-ray spectrum of
\novamus\ \citep{Ebisawa:1994a}.  \citet{Misra:1999a} did address this
question in the context of their spectral modeling and show that the
two are consistent provided that Compton scattering modifies the X-ray
spectrum with a spectral hardening factor \citep{Shimura:1995a} of
$f_c \sim 3$.  This factor is larger than the value of 1.7 usually
adopted, but \citet{Merloni:2000a} argued that values as large as 3
are plausible.  If one accepts this correction then both the
optical/UV spectrum, and the soft X-ray thermal component can be
consistently interpreted with a purely viscously heated disk.
As noted above for \novamon, irradiation may still play a role
in stabilizing the disk in the high viscosity state, but it is not
required to provide the main energy input.  We will consider an
alternative explanation for the optical/UV spectrum in
Section~\ref{CoronaSection}.

\subsection{\novaper}
For \novaper\ we have two useful epochs with both \IUE\ and optical
data.  These data may also suffer from similar problems with the
far-UV rise component of extinction to \novamus\ and \novaper, but in
the opposite sense, as there seems to be an enhanced drop off in flux
at the shortest wavelengths.  

\begin{figure}
\includegraphics[angle=90,scale=0.37]{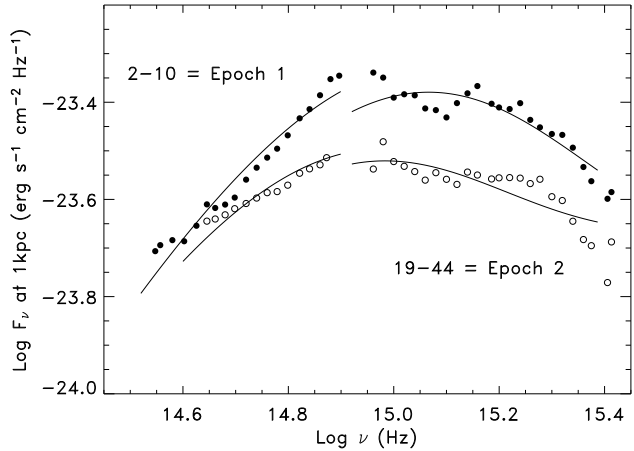}
\caption{Disk model fits to the spectra of \novaper.  Offsets have
  been included to allow for non-simultaneity, but only a small
  difference is required.}
\label{NPerFitFig}
\end{figure}

Fits to these data are shown in Fig.~\ref{NPerFitFig}.  Strong
irradiative heating is needed in this case, although as the outburst
declines viscous heating may be becoming important as in \novavulb;
the timebase and data quality are inadequate to say this with any
confidence, however.  The irradiation temperatures inferred are lower
than for \novavulb, with $T_{\rm irr} \sim 13\,000$\,K early in
outburst, compared to $\sim 18\,000$\,K in \novavulb\
\citep{Hynes:2002a}.  Lower temperatures are consistent with the
softer optical spectrum earlier estimated.  As seen in
Fig.~\ref{UVSEDFig}, this difference does seem to be reflected in the
lower peak frequency of the SED, so is likely real.  As for the other
sources, we can estimate the disk radius required; assuming
$d=2.49\pm0.30$\,kpc and $i=45\pm2^{\circ}$ \citep{Gelino:2003a}, we
derived disk radii of $(1.3\pm0.3)\times10^{11}$\,cm and
$(1.7\pm0.5)\times10^{11}$\,cm for the first and second epochs
respectively.  These numbers seem too large, as we estimate a tidal
truncation radius of $\sim9\times10^{10}$\,cm assuming the parameters
of \citet{Gelino:2003a}.  This discrepancy could be resolved with a
reduction in the source distance (which is not totally impossible;
\citealt{Webb:2000a}), with a more modest increase in the
temperature (which is factored into the large uncertainty in this
derived radius), or if the inclination uncertainty is larger
  than assumed.
\subsection{Indirect heating of the disk}
\label{CoronaSection}
We have seen that UV-hard spectra are ubiquitous, and can occur even when
a BHXRT is at a high optical luminosity.  It is common, at least in
the field of X-ray binaries, to interpret such a spectrum as a
signature of a viscously heated steady-state disk (e.g.\
\citealt{Cheng:1992a}), and thus far we have done the same.  This is
not implausible, particularly in the case of \novamon\ for which the
required mass flow rate is not large.

All this spectral form really indicates, however, is the temperature
distribution, and that the heating rate varies as approximately
$R^{-3}$.  An alternative irradiation model is used by the young star
community which actually predicts irradiative heating varying as
$R^{-3}$ (e.g.\ \citealt{Hartmann:1998a}).  This model addresses the
case where a circumstellar disk is irradiated by the stellar surface.
The key difference is that since the star is not compact, it is not a
point source, but an extended one which illuminates the disk from
above.  Indeed a similar case which also produces $R^{-3}$ heating
arises from a `lamppost' model, where the source of X-rays are raised
above the disk (e.g.\ \citealt{Collin:1990a}).

As described in the appendix, one can produce UV-hard spectra with an
elevated (or vertically extended) X-ray source.  This behavior is
generically expected, except for the specific case of a concave disk
with a sufficiently compact X-ray source, possibly mediated by local
scattering by an extended disk atmosphere.  It is worth emphasizing
that where irradiation is attributed to X-rays scattered by a central
coronal source, one {\em expects} $R^{-3}$ heating, in contrast to the
assumption usually made in modeling outburst behavior that irradiation
drops off approximately as $R^{-2}$ (e.g.\ \citealt{King:1998a},
\citealt{Kim:1999a}, \citealt{Esin:2000a}, \citealt{Dubus:2001a}).  If
X-ray scattering occurs in a highly vertically extended disk
atmosphere, however, $R^{-2}$ heating could still be appropriate.

How plausible are these scenarios?  X-ray scattering `coronae' are
certainly believed to be present in X-ray binaries.  However, it is
likely that coronae are not spherical, but very flattened, and
represent a vertical extension of the disk atmosphere (e.g.\
\citealt{Church:2003a}).  Effective illumination of a convex disk,
with a vertically extended central corona, thus seems difficult to
achieve, and hence this seems unlikely as a mechanism for producing
UV-hard spectra.

Finally, we note that X-rays from a jet might provide an alternative
`lamppost' illumination, either by direct synchrotron emission or by
scattering.  Relativistic beaming might then be expected to
considerably reduce the back-illumination of the disk, but large
luminosities are not required, especially as the X-ray source could
then be higher above the disk than is plausible for a corona.
\subsection{Heating of a warped disk}
\label{WarpSection}
\citet{Dubus:1999a} suggested that warping of disks may facilitate
irradiation of disks which would otherwise be self-shielded, and this
could also be expected to modify the radial dependence of the heating.
\citet{Ogilvie:2001a} have examined this question quantitatively.
They do find strong variations in the radially averaged heating, but
these are not as simple as the additional $1/R$ dependence that would
be required to reproduce the $F_{\nu}\propto \nu^{1/3}$ spectra,
although they do find that an average reprocessing efficiency of
$C=5\times10^{-3}$ can readily be produced with a warped disk.  While
only a fraction of the disk is illuminated at any time, the effect of
irradiation on the thermal structure of the disk will last for of
order the local thermal timescale.  This should be comparable to the
local dynamical timescale (i.e.\ the Keplerian rotation period), so
heating effects will be azimuthally averaged as material rotates
through the illuminated patch and hence the whole disk will be
affected (G. Dubus, priv.\ comm.).  Further investigation of the
spectral energy distributions expected in the warped disk case is
needed.  \citet{Ogilvie:2001a} do note, however, that the short-period
BHXRTs we are considering here are expected to be stable against
radiation driven warping, so in the absence of another warping
mechanism, it remains questionable whether warping can be an important
factor for these objects.

%
\section{Discussion}
\label{DiscussionSection}
In light of the above discussion, we should reconsider the origin of
the optical emission in BHXRTs in outburst.  It is sometimes stated
that optical emission is always totally dominated by reprocessed
X-rays.  This may be true in neutron star systems, but there is good
reason to {\em expect} irradiation to be weaker in the black hole
case, as there is no direct emission from the compact object surface
\citep{King:1997a}.  This difference is invoked by these authors to
explain the observations that black hole LMXBs are transient, whereas
most analogous neutron star systems are persistent X-ray sources.

The studies presented here suggest that the situation may be more
marginal than usually assumed.  If one interprets a $\nu^{1/3}$
spectrum as a signature of a viscously heated disk, then the answer is
clearly no, as this spectrum occurs rather often among the population.
In the case of \novamon, at least, if one takes a plausible mass flow
rate, and plausible disk size then viscous heating alone can readily
explain the observed optical brightness.  In the case of \novavulb,
the SED can initially be interpreted as irradiation dominated, but a
transition to the viscously dominated regime appears to occur
suggesting that the irradiation dominance is rather marginal.  Of
course this conclusion hinges upon one interpretation of the SED, but
the alternatives considered are relatively unpalatable.  Irradiation
by X-rays from a lamppost (perhaps the jet?), or by a vertically
extended corona could reproduce the observed SED, but requires an
excessive luminosity for the extended source and/or a rather large
vertical extent.  We thus appear left with viscous heating of the
outer disk as the most natural explanation of the observed optical
flux during outburst, at least when UV-hard spectra are seen.  As
noted several times, irradiation should still play a role in
maintaining the disk in a hot, high viscosity state by suppressing the
disk instability, but it does not appear to dominate the energy
budget.

The point is often made that studies such as this one are severely
compromised by assumption of local black-body spectra.  At some level,
of course, this is true, but in the UV, and especially the optical,
the effect is likely to be mild.  The two most important effects on a
optical-UV broad-band SED, for the temperatures considered, are
hydrogen breaks and limb-darkening.  Line-blanketing by many blended
absorption lines could have some effect, although the highest quality
data suggest a rather clean continuum where the effect of intrinsic
(non-interstellar) absorption lines is very weak or absent, except for
Ly$\alpha$ \citep{Haswell:2002a}.  The Balmer jump falls well within
our spectral coverage.  While this results in strong distortion of the
SED of hot stars, such large effects are clearly not present in our
data, and there is little if any evidence for its presence at all.
The observations are consistent with spectral models of accretion
disks around stellar mass black holes which do predict rather weak
Balmer jumps (e.g.\ \citealt{Hubeny:2001a}; Hubeny priv.\ comm.)  A
much larger effect is predicted at the Lyman break, and has the effect
of significantly enhancing the far-UV flux longward of the break.
This effect produces an upturn which is principle is detectable.  Such
an upturn was indeed observed in \novavulb\ \citep{Hynes:2002a}, and
may also be present in \novamus, as shown earlier.  An alternative
explanation was advanced by \citet{Hynes:2002a}, however, that this upturn
originates from deviation of the far-UV rise component of extinction
from the Galactic average.  Disentangling the two effects is likely
impossible without line-of-sight extinction curves or a larger sample
of objects.  Furthermore, strong Lyman breaks can also be predicted in
AGN spectra, but are not observed (see \citealt{Blaes:2001a} and
references therein).  The other deviation from black body spectra that
could be significant is limb-darkening \citep{Diaz:1996a}.  This is
wavelength dependent and stronger in the UV, so tends to soften the
far-UV flux.  It is an inclination dependent effect, however, and
for the moderate inclinations of the objects considered here
(40--60$^{\circ}$; see Section~\ref{ModelSection}) should not be
dominant.  With both Balmer and Lyman breaks and limb-darkening, the
models of \citet{Hubeny:2001a} do not deviate from their black-body
counterparts by more than $\pm0.1$ in $\log F_{\nu}$ over the
optical-UV range, for inclinations of less than 60$^{\circ}$.  In
particular, a UV-hard spectrum remains UV-hard.  Consequently
black-body spectra do appear adequate for the analysis presented here,
and it is likely that the dominant uncertainty is actually in the
reddening corrections, not the local emergent spectra.

%
\section{Conclusions}
\label{ConclusionSection}
The optical and UV spectral energy distributions of BHXRTs in outburst
have the potential to inform us of the temperature distribution of the
outer disk, and hence test models for the heating of these disks.  We
have collated all the spectra published to date 
on short period typical systems which we believe are useful in this
regard.  In spite of a 25\,year baseline, these data are rather
limited.  \HST\ data are by far the most suitable for this work in
both the UV and the optical.  Ground-based optical studies suffer from
uncertainties in the flux calibration, especially at short
wavelengths, and from Telluric absorption at long wavelengths.
Consequently even from the small sample available, only a subset are
really reliable.

Even with high quality data, further complications arise from
uncertainty in the line-of-sight extinction curve.  If the curve is
known, then the reddening can be determined to high-precision using
the 2175\,\AA\ feature.  Without {\it a priori} knowledge of the
extinction curve, however, the relative strength of the feature is not
known and the reddening cannot be precisely determined.  Further
difficulties arise from variations in other aspects of the UV
extinction curve.  The far-UV rise seems particularly variable in our
sample and several of the sources considered exhibited kinks in the
far-UV which could be attributed to an incorrect correction of the
far-UV rise.  These further compound the uncertainty in the dereddened
spectral energy distributions.

In spite of these difficulties, we can draw some useful conclusions.
Provided a suitable sample of typical short-period BHXRTs are
selected, the optical spectra form a relatively homogeneous sample.
All have quasi-power-law shapes, with spectral slopes (in $F_{\nu}$
vs.\ $\nu$) of 0.5--1.5.  Where the data are good enough to
discriminate, the optical spectra tend to soften as the outburst
decays, corresponding to the cooling of the outer disk.  The very
homogeneity of the optical spectra means that they are of limited
value alone in testing disk models, and they cannot discriminate
between different forms of disk heating.  It is also worth remarking
that since the optical lies on the long wavelength tail of the
spectral energy distribution the $V$ band flux is not necessarily
related in a straightforward way to the irradiating luminosity and
this may complicate attempts to compare optical and X-ray lightcurves;
for example the optical brightness {\em at a fixed mass flow rate},
and hence the optical to X-ray flux ratio, will depend upon the size
of the hot region of the disk.

To perform more useful tests requires UV data.  Ideally full optical-UV
coverage can be used, but the UV alone has some power to
discriminate.  All spectra flatten in the UV, although depending on
the source and epoch this effect may be a modest reduction in slope, or a
complete turnover (i.e.\ decreasing $F_{\nu}$).  Uncertainties in
dereddening the spectra complicate the interpretation of the UV slope,
but in \novavulb\ evolution was seen from one form to the other,
indicating that there are real variations.  In this case, the
evolution was in the opposite sense to the optical; the UV spectrum
hardened during the outburst decay.  These spectral shapes can be
thought of as defining optical-UV spectral states in BHXRTs, analogous
to X-ray spectral states.  The case where the UV spectrum continues to
rise can be termed the UV-hard state, and both \novamon\ and \novamus\
exhibit this state.  Spectra which peak and fall off in the UV can be
termed UV-soft, such as those of \novaper.  In this terminology,
\novavulb\ exhibited a transition from the UV-soft to UV-hard state as
it decayed.

These spectra can be interpreted in terms of simple black body disk
models.  The full range can be accommodated with two forms of heating,
one varying as $R^{-2}$ and one as $R^{-3}$.  The first is readily
interpreted as due to irradiation by a compact X-ray source and
produces UV-soft spectra.  The second is usually associated with
viscous heating, but can also arise if the disk is illuminated by a
vertically extended X-ray source.  Evolution from the first form to
the second as the disk cools results in exactly the behavior seen in
\novavulb\ -- the optical spectrum softens, but the UV hardens.  None
of the other sources exhibit such a clear transition, although
\novaper\ exhibited similar optical spectral evolution to \novavulb\
suggested it might have shown the same effect if suitable late-epoch
data were available.  \novamon\ and \novamus\ suggest $R^{-3}$
heating, whereas \novaper\ can be interpreted as $R^{-2}$ irradiated
heating.  It might be significant that both \novaper\ and \novavulb\
(at least early in the outburst) were identified by
\citet{Brocksopp:2002a} as particularly hard sources; it may be that
there is a correlation between the optical-UV state (i.e.\ the form of
heating of the outer disk) and the X-ray state.  Evidence for this
effect is provided by \citet{Esin:2000a} who argue that the transition
to the low/hard state near the end of the outburst is accompanied by a
substantial enhancement in the reprocessing efficiency.  Obviously a
larger sample is needed to robustly test these issues.  Ideally we
would like to see more examples of transitions in the optical-UV
spectral state.

The interpretation of the UV-hard state remains uncertain.  These
spectra are well fitted by a viscously heated Shakura-Sunyaev disk
spectrum.  In the case of \novamon\ this interpretation is quite
sensible -- assuming a lobe-filling disk accreting at 30\,\%\ of the
Eddington rate quite naturally produces both the optical fluxes and
the spectral shape without invoking reprocessing as the dominant
source of optical emission.  To explain \novamus\ in the same way,
however, appears to require super-Eddington accretion (c.f.\
\citealt{Misra:1999a}).  Irradiation by an extended X-ray source
offers a way out of this difficulty, and could bypass self-shielding
of the disk \citep{Dubus:1999a}, but creates its own problems by
requiring a very extended X-ray emission (or scattering) region.

Further progress in this area undoubtedly requires more observations
of the quality of those of \novavulb.  Just as important, however, may
be for the observations to make contact meaningfully with simulations
of disk outbursts.  Progress is being made (e.g.\
\citealt{Lasota:2001a}) but simulations have not yet adequately
accounted for disk warping, or realistic X-ray irradiation
\citep{Dubus:1999a}; the X-ray emission geometry (compact vs.\
extended), and spectral changes may have a significant impact on the
strength of irradiation (e.g.\ \citealt{Esin:2000a}).  
%

\acknowledgments

I am grateful for funding from NASA through Hubble Fellowship grant
\#HF-01150.01-A awarded by STScI, which is operated by the AURA, Inc.,
for NASA, under contract NAS 5-26555.  I would like to thank Chris
Shrader, Nicola Masetti, and Massimo Della Valle for providing digital
optical spectra presented here; Dawn Gelino for many of the distances;
Erik Kuulkers for helpful discussions on \novamon; and Ivan Hubeny and
Richard Wade for discussions of disk spectral modeling.  Particular
thanks are due to Carole Haswell for originally prompting me to
investigate this issue, much fruitful collaboration in this and
related areas, and specifically on the data on \novavulb\ included
here, and to Rob Robinson for careful reading of the manuscript.
Finally I am grateful to the referee, Guillaume Dubus, for
constructive criticism and correction on a number of points.

Some of the data included in this paper were obtained from the
Multimission Archive at the Space Telescope Science Institute (MAST)
or the ESO/ST-ECF Archive Facility.  STScI is operated by the
Association of Universities for Research in Astronomy, Inc.\ under
NASA contract NAS5-26555.  Specifically, this work includes
observations made with the NASA/ESA Hubble Space Telescope and the
NASA/ESA/SERC International Ultraviolet Explorer.  Support for MAST
for non-HST data is provided by the NASA Office of Space Science via
grant NAG-7584 and by other grants and contracts.  \IUE\ data were
obtained under proposals XBMCS (\novamus, PI Shrader), MATOO, MITOO,
N1205 (\novamus) and OD45Z, XROCS (\novaper, PI Shrader).
\HST\ data were obtained under proposals 3232 (\novamus, PI Panagia)
and 8245 (\novavulb, PI Haswell).

This work has made extensive use of the NASA
Astrophysics Data System Abstract Service. 

%
\appendix
\section{Irradiation from a central scattering corona}

We can explore different irradiation regimes qualitatively with a simple
geometric model of reprocessing, similar to that used by
\citet{Collin:1990a} for active galactic nuclei.  Consider a disk
element at radius $R$, height $H$, and local slope ${\rm d}H/{\rm
d}R$.  The element is illuminated by an isotropically emitting X-ray
point source at height $Z$ above the center of the disk.  The flux $F$
incident upon the surface element depends upon both the distance from
the X-ray source and the angle of incidence of the X-rays.  In
general,
\begin{equation}
F \propto \frac{\hat{\bm n}\cdot \hat{\bm d}}{\left|{\bm d}\right|^2}
\end{equation}
where the distance of the X-ray source from the element is ${\bm
d}=(-R,Z-H)$ in cylindrical coordinates, and the normal to the surface
is $\hat{\bm n}=\left( 1 + ({\rm d}H/{\rm d}R)^2 \right)^{-1/2} \left(
-{\rm d}H/{\rm d}R, 1\right)$.  For a geometrically thin disk ($H \ll
R$), compact X-ray source ($Z \ll R$), and shallow disk gradients
($dH/dR \ll 1$), this result simplifies to:
\begin{equation}
F \propto \frac{R \left({\rm d}H/{\rm d}R\right) + (Z-H)}{R^3}
\end{equation}

The simplest geometry is a flat disk ($H={\rm d}H/{\rm d}R=0$).  In
this case, $F\propto Z/R^3$; this is the simple `lamppost' case
referred to above.  The same would apply if the disk is not flat, but
${\rm d}H/{\rm d}R$ is constant, since then $H=R{\rm d}H/{\rm d}R$.
Somewhat more realistic cases can be approximated with a power-law
disk: $H=H_0 (R/R_0)^{\beta}$, where $H_0$ and $R_0$ refer to values
at the outer disk edge.  In this case,
\begin{equation}
F \propto \frac{\left( \beta-1 \right)H + Z}{R^3}
\end{equation}

$\beta=1$ is the case described above, and separates the behavior into
two regimes.  Firstly consider a concave disk, $\beta>1$.  If $\left(
\beta-1 \right)H \ll Z$ then the curvature of the disk is negligible
and $F\propto Z/R^3$ as for a flat disk.  If $\left( \beta-1 \right)H
\gg Z$ then the vertical elevation of the X-ray source is negligible
and we obtain the classical behavior of a concave disk illuminated by
a central point source, with $F\propto R^{-2}$ in the limiting case of
infinitesimal curvature (i.e.\ $\beta-1 \ll 1$).  The dividing
line between the two thus depends upon both the height of the disk
relative to the height of the X-ray source and the amount of curvature
of the disk.  For an extended X-ray source, radius $Z$, we might then
expect a UV-hard spectrum (corresponding to $F\propto R^{-3}$) if the
X-ray source has size larger than $\left( \beta-1 \right)H$.  

Next consider a convex disk, $\beta < 1$.  In this case one {\em
never} expects heating to vary as $R^{-2}$.  If $\left(1- \beta
\right)H > Z$ then the X-ray source is below the local horizon of the
surface element and no irradiation is possible.  If $\left(1- \beta
\right)H \ll Z$ then again disk curvature is negligible and $F\propto
Z/R^3$.  In the intermediate cases the radial dependence is even
steeper.  The caveat to this occurs if there is also local scattering
material above the disk (i.e.\ the disk atmosphere) which can scatter
irradiation over the disk horizon and illuminate regions that would
otherwise be obscured.  In this case, $R^{-2}$ heating remains possible.

%





\end{document}